\documentclass[a4paper,final,oneside,notitlepage,12pt]{article}

\usepackage[T1]{fontenc}
\usepackage{amsfonts}
\usepackage{amssymb}
\usepackage{amsthm}
\usepackage{amsmath}
\usepackage[all]{xy}
\usepackage{amsfonts}
\usepackage{amssymb}
\usepackage{amsmath}
\usepackage{color}

\usepackage[section]{placeins}

\usepackage{graphicx}

\newtheoremstyle{mythmst}{\topsep}{\topsep}{\it}{\parindent}{}{\@addpunct{.}}{ }{\sc{#1 #2}\if#3\else\hspace{1mm}\normalfont (#3)\fi}  
\newtheorem{definition}{Definition}
\newtheorem{remark}{Remark}

\newtheorem{lemma}{Lemma}
\newtheorem{theorem}{Theorem}

\def\maps{\;\;:\;\;}
\def\lto{\;\;\longrightarrow\;\;}
\def\Lto{\;\;\Longrightarrow\;\;}
\def\mapslto{\;\;\longmapsto\;\;}

\def\Z {\mathbb{Z}}

\def\R {\mathbb{R}}
\def\C {\mathbb{C}}
\def\im{\mathrm{i}}
\def\Id {\mathrm{id}}
\def\id {\mathrm{id}}

\def\worldsheet {\Sigma}
\def\worldsheetCoveringSpace {\hat\Sigma}
\def\worldsheetCover {\mathcal{U}}
\def\worldsheetCovering {\mathrm{pr}}
\def\worldsheetInvolution{\sigma}

\def\orbifoldGroup {K}
\def\orbifoldGroupElement {k}

\def\targetspace{M}
\def\targetspaceCover{V}
\def\targetspaceNerve{\targetspace_{\mathfrak{\targetspaceCover}}}

\def\gerbe{\mathcal{G}}
\def\gerbeloci{B}
\def\gerbelocij{A}
\def\gerbelocijk{g}

\def\jandl{\mathcal{J}}
\def\jandlloci{W}
\def\jandllocij{t}

\def\deligneDifferential{\mathrm{D}}
\def\exteriorDifferential{\mathrm{d}}
\def\dlog{\mathrm{dlog}}
\def\gerbetrivi{M}
\def\gerbetrivij{h}
\def\stringmap{\phi}
\def\jandlloccc{j}
\def\worldsheetCover{U}

\makeatletter

\long\def\@makecaption#1#2{
  \vskip\abovecaptionskip
  \sbox\@tempboxa{{\bf #1}: {#2}}
  \ifdim \wd\@tempboxa >10cm
    \begin{list}{}{
      \leftmargin=15mm
      \labelwidth=0mm
      \rightmargin=15mm
      \listparindent=0mm
      \itemsep=0mm
    }  
    \item{\bf #1}: {#2}\end{list}
  \else
    \global \@minipagefalse
    \hb@xt@\hsize{\hfil\box\@tempboxa\hfil}
  \fi
  \vskip\belowcaptionskip}

\def\proof {{Proof.}\hspace{7pt}}
\def\endofproof {\hfill{$\square$}\\}

\def\adress#1{\gdef\@adress{#1}}
\def\@adress{}
\def\preprint#1{\gdef\@preprint{#1}}
\def\@preprint{}
\def\@maketitle{
  \newpage
  \noindent
  \begin{tabular}{cc}
    \begin{minipage}[c]{0.4\textwidth}
      \begin{flushleft}
        \includegraphics[width=110pt]{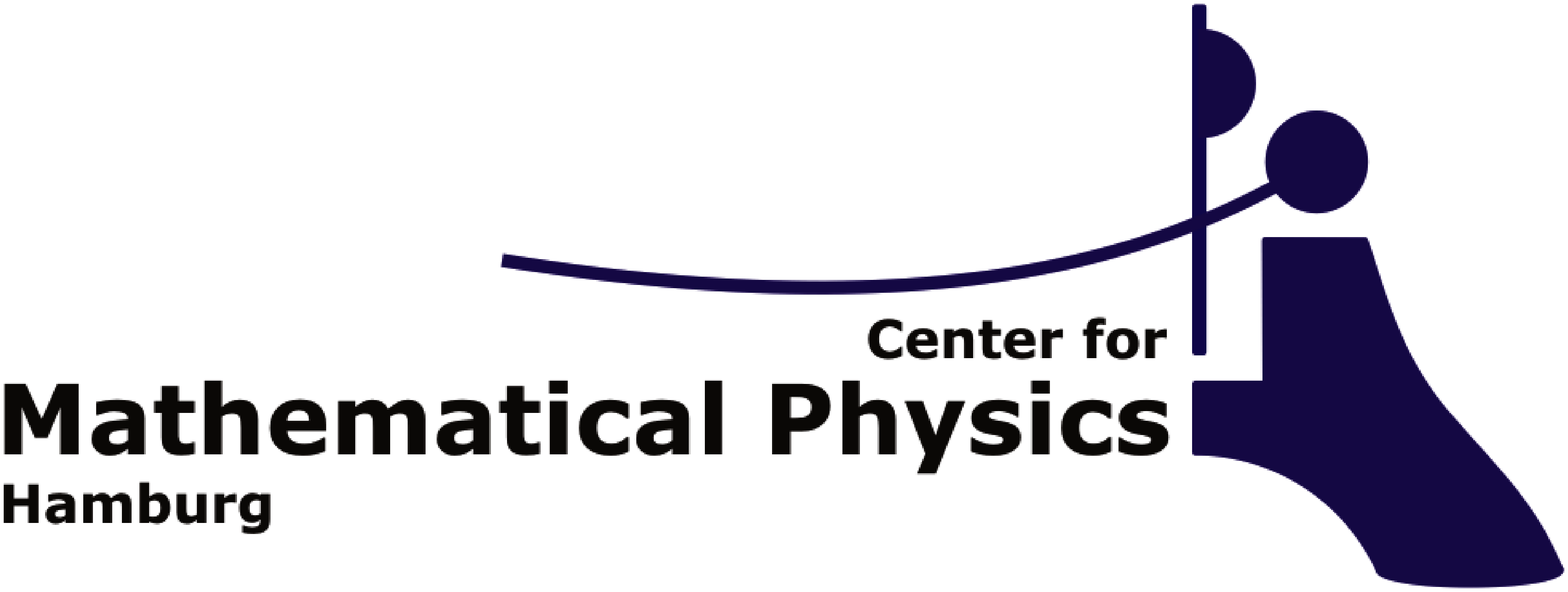}
      \end{flushleft}  
    \end{minipage}&
    \begin{minipage}[c]{0.6\textwidth}
      \begin{flushright}
      {\small\sf\@preprint}
      \end{flushright}
    \end{minipage}
  \end{tabular}
  \vskip 3cm
  \begin{center}
    \LARGE\@title
    \if!\@author!\else \vskip 0.5cm \large\@author\fi
    \if!\@adress!\else \vskip 0.5cm \normalsize\@adress\fi
  \end{center}
  \vskip 2cm
}

\makeatother

\begin{document}

\title{Unoriented WZW Models and \\ Holonomy of Bundle Gerbes}

\author{Urs Schreiber$^1$,
 Christoph Schweigert$^2$,
 Konrad Waldorf$^3$}

\adress{Fachbereich Mathematik\\Schwerpunkt Algebra und Zahlentheorie\\Universit\"at
Hamburg\\Bundesstra\ss e 55\\D--20146 Hamburg
}

\preprint{arXiv:hep-th/0512283\\
Hamburger Beiträge zur Mathematik Nr. 228\\
ZMP-HH/05-28}

\maketitle

\begin{abstract}
\noindent The Wess-Zumino term in two-dimensional conformal field theory is best understood
as a surface holonomy of a bundle gerbe. We define additional structure for a bundle
gerbe that allows to extend  the notion of surface holonomy to unoriented
surfaces. This provides a candidate for the Wess-Zumino term for WZW models
on unoriented surfaces. Our ansatz reproduces some results known from the
algebraic approach to WZW models.
\end{abstract}

\thispagestyle{empty}

\footnotetext[1]{Email: schreiber@math.uni-hamburg.de.}
\footnotetext[2]{Email: schweigert@math.uni-hamburg.de.}
\footnotetext[3]{Email: konrad.waldorf@math.uni-hamburg.de.
 K.W. is supported with scholarships
by the German Israeli Foundation (GIF) and
by the Rudolf und Erika Koch--Stiftung.}

\newpage
\thispagestyle{empty}

\noindent manche meinen\\
lechts und rinks\\
kann man nicht velwechsern\\
werch ein illtum

\vspace{0.5cm}
\hspace{2cm}Ernst Jandl \cite{jandl1}

\vfill
\setcounter{tocdepth}{2}
\tableofcontents

\newpage
\section{Introduction}

Wess-Zumino-Witten (WZW) models are one of the most important classes of 
(two-dimensional) rational conformal field theories. They describe physical 
systems with (non-abelian) current symmetries, provide gauge sectors in
heterotic string compactifications and are the starting point for other
constructions of conformal field theories, e.g.\ the coset construction.
Moreover, they have played a crucial role as a bridge between
Lie theory and conformal field theory.

It is well-known that for the Lagrangian description of such a model,
a Wess-Zumino term is needed to get a conformally invariant theory
\cite{witten1}. Later, the relation of this term to Deligne hypercohomology has
been realized  \cite{gawedzki3} and its nature as a surface holonomy
has been identified \cite{gawedzki3,alvarez1}. More recently, the
appropriate differential-geometric object for the holonomy has been identified 
as a hermitian $U(1)$ bundle gerbe with connection and curving \cite{carey2}. 

Already the case of non-simply connected Lie groups with non--cyclic fundamental
group, such as $G:={Spin}(4n) / \Z_2\times \Z_2$ shows that gerbes and
their holonomy are really indispensable, even when one restricts one's attention
to oriented surfaces without boundary. The original definition of the 
Wess-Zumino term as the integral of a three form $H$ over a suitable
three-manifold cannot
be applied to such groups; moreover, it could not explain the well-established 
fact that to such a group {\em two} different rational conformal field theories
that differ by \textquotedblleft discrete torsion\textquotedblright\ can be associated.

Bundle gerbes will be central for the problem we address in this paper. A 
long series of algebraic results indicate that the WZW model can be
consistently considered on unorientable surfaces. Early results include
a detailed study of the abelian case \cite{bianchi1} and of ${ SU}(2)$
\cite{pradisi2,pradisi3}. Sewing constraints for unoriented surfaces have been 
derived in \cite{fioravanti1}.

Already the abelian case \cite{bianchi1} shows 
that not every rational conformal field theory that is well-defined
on oriented surfaces can be considered on unoriented surfaces. A necessary
condition is that the bulk partition function is symmetric under exchange
of left and right movers. This restricts, for example, the values
of the Kalb-Ramond field in toroidal compactifications \cite{bianchi1}.
Moreover, if the theory can be extended to unoriented surfaces, there can be 
different extensions that yield inequivalent correlation functions.
This has been studied in detail for WZW theories based on ${SU}(2)$ in 
\cite{pradisi2,pradisi3}; later on, this has been systematically described with
simple current techniques \cite{huiszoon2,huiszoon3}. Unifying general formulae have
been proposed in \cite{fuchs1}; the structure has been studied at the
level of NIMreps in \cite{sousa1}. Aspects of these results have been proven
in \cite{fuchs2} combining topological field theory in three-dimensions
with algebra and representation theory in modular tensor categories.
As a crucial ingredient, a generalization of the notion of an algebra
with involution, i.e. an algebra together with an algebra-isomorphism
to the opposed algebra, has been identified in \cite{fuchs2}; the isomorphism
is not an involution any longer, but squares to the twist on the
algebra. An algebra with such an isomorphism has been called 
Jandl algebra. A similar structure, in a geometric setting, will be the
subject of the present article.

\medskip

The success of the algebraic theory leads, in the Lagrangian description, to the 
quest for corresponding geometric structures on the target space. From previous 
work \cite{bachas1,huiszoon1,brunner1} it is clear that a map $k\,: M \to M$ 
on the target space with the additional property that $k^*H=-H$ will be one 
ingredient. 
Examples like the Lie group ${SO}(3)$, for which two different unoriented WZW models with
the same map $k$ are known, already show that this 
structure does not suffice.

We are thus looking for an additional structure on a hermitian bundle gerbe 
which allows to define a Wess-Zumino term, i.e.\ which allows to define
holonomy for unoriented surfaces. For a general bundle gerbe, such a 
structure need not exist; if it exists, it will not be unique. 

In the present article, we make a proposal for such a structure.
It exists whenever there are sufficiently well-behaved stable isomorphisms
between the pullback gerbe $k^*{\mathcal G}$ and the dual gerbe
$\mathcal G^{*}$. If one thinks about a gerbe as a sheaf of groupoids,
the formal similarity to the Jandl structures in \cite{fuchs2} becomes
apparent, if one realizes that the dual gerbe plays the role of the
opposed algebra. For this reason, we term the relevant structure a Jandl 
structure
on the gerbe. We show that the Jandl structures on a gerbe on the target
space $M$, if they exist at all, form a torsor over the group of flat 
equivariant hermitian line bundles on $M$. As explained in section 4.3,
this group always contains an element $L^k_{-1}$ of order two. We show that
two Jandl structures that are related by the action of $L^k_{-1}$ provide
amplitudes that just differ by a sign that depends only on the topology
of the worldsheet. Such Jandl structures are considered to be essentially
equivalent. We finally show that 
a Jandl structure allows to extend the definition of the usual
gerbe holonomy from oriented surfaces to unoriented surfaces.
We derive formulae for these holonomies
in local data that generalize the formulae of \cite{gawedzki1,alvarez1} for
oriented surfaces.

\medskip

To give a concrete impression of a Jandl structure,
we write out the local data of a Jandl
structure for a given gerbe $\mathcal G$ on the target space $M$.
To this end, we first recall the local data of a hermitian
bundle gerbe in a good open cover $\{\targetspaceCover_i\}_{i\in I}$ of 
$M$: we have a 2-form $B_i$ for each open set $V_i$, a 1-form 
$A_{ij}$ on each intersection $V_i
\cap V_j$ and a $U(1)$-valued
function $g_{ijk}$ on each triple intersection $V_i\cap V_j\cap V_k$.
They are required to satisfy the following constraints:
\begin{eqnarray*}
\gerbelocijk_{jkl}\cdot \gerbelocijk_{ikl}^{-1}\cdot \gerbelocijk_{ijl}\cdot \gerbelocijk_{ijk}^{-1} &=&1
\\
\gerbelocij_{jk}-\gerbelocij_{ik}+\gerbelocij_{ij}+\dlog\left( \gerbelocijk_{ijk}\right) &=&0
\\
-\exteriorDifferential \gerbelocij_{ij}+\gerbeloci_{j}-\gerbeloci_{i} &=&0\text{.}   
\end{eqnarray*}

To write down the local data of a Jandl structure for a given involution
$k:M\to M$ in a succinct manner, we make the simplifying assumption that we 
have a cover $\{V_i\}_{i\in I}$ that is invariant under $k$, $k(V_i)=V_i$, and 
that is still good enough to provide local data. The local data of a Jandl
structure then consist of a $U(1)$-valued function $j_i\,: V_i\to U(1)$
for each open subset, a $U(1)$-valued function $t_{ij}\,: V_i\cap V_j
\to U(1)$ on two-fold intersections and a 1-form $W_i\in\Omega^1(V_i)$.

They relate the pullbacks of the gerbe data under $k$ to the local data of the dual gerbe
as follows:
\begin{eqnarray*}
\orbifoldGroupElement^{*}\gerbeloci_{i} &=& - \gerbeloci_{i}
+\exteriorDifferential \jandlloci_{i} \\
\orbifoldGroupElement^{*} \gerbelocij_{ij} &=& - \gerbelocij_{ij}
 -\dlog(\jandllocij_{ij})
+ \jandlloci_{j} - \jandlloci_{i} \\
\orbifoldGroupElement^{*}\gerbelocijk_{ijk} &=& \gerbelocijk_{ijk}^{-1}
\cdot t_{jk} \cdot t_{ik}^{-1} \cdot t_{ij} 
\end{eqnarray*}
The local data of a Jandl structure are required
to be equivariant under  $\orbifoldGroupElement$
in the sense that
\begin{eqnarray*}
\orbifoldGroupElement ^{*} \jandlloci _{i} &=& \jandlloci _{i} - \dlog (\jandlloccc_{i}) \\
\orbifoldGroupElement ^{*}\jandllocij_{ij}  &=&  \jandllocij_{ij} \cdot \jandlloccc_{j}^{-1}
 \cdot \jandlloccc_{i} \\
\orbifoldGroupElement^{*}\jandlloccc_i &=& \jandlloccc_i^{-1}\text{.}\end{eqnarray*}
It should be appreciated that the functions $t_{ij}$
are {\em not} transition functions of some line bundle; as we will explain in
section 2.4, they are rather the local data describing an isomorphism of
line bundles appearing in the Jandl structure.

The notion of a Jandl structure naturally explains algebraic results
for specific classes of rational conformal field theories. It is well-known
that both the Lie group $SU(2)$ and its quotient $SO(3)$ admit two
Jandl structures that are essentially different (i.e.\ that do not just
differ by a sign depending on the topology of the surface). In the
case of $SU(2)$, this is explained by the fact that two different 
involutions are relevant: $g \mapsto g^{-1}$ and $g\mapsto z g^{-1}$,
where $z$ is the non-trivial element in the center of $SU(2)$. Indeed,
since $SU(2)$ is simply-connected, we have a single flat line bundle
and hence for each involution only two Jandl structures which are essentially
the same.

The two involutions
of $SU(2)$ descend to the same involution of the quotient $SO(3)$.
The latter manifold, however, has fundamental group $\Z_2$ and thus
twice as many equivariant flat line bundles as $SU(2)$. The different
Jandl structures of $SO(3)$ are therefore not explained by different
involutions on the target space but rather by the fact that one involution
admits two essentially different Jandl structures.

\medskip

Needless to say, there remain many open questions. A discussion of surfaces
with boundaries is beyond the scope of this article. The results of
\cite{fuchs2} suggest, however, that a Jandl structure leads to an involution
on gerbe modules. Most importantly,
it remains to be shown that, in the Wess-Zumino-Witten path integral 
for a surface $\worldsheet$, the holonomy we introduced yields amplitudes that take 
their values in the space of conformal blocks associated to the complex
double of $\worldsheet$, which ensures that the relevant chiral Ward identities are
obeyed. To this end, it will be important to have a suitable reformulation
of Jandl structures at our disposal. Indeed, the holonomy we propose
in this article also arises as the surface holonomy of a 2-vector bundle
with a certain 2-group; this issue will be the subject of a separate
publication.

\section{Bundle Gerbes with Jandl Structures}

\subsection{Bundle Gerbes and stable Isomorphisms}
\label{ss_bg_bg}

In preparation of the following sections,
in this section we define an equivalence relation
on the set of stable isomorphisms between two fixed bundle gerbes. To this end, we first set up the notation concerning
bundle gerbes and stable isomorphisms. We  mainly adopt   the formalism
used by Murray and collaborators, see \cite{carey2} for example, as well as by  Gaw\c{e}dzki and Reis \cite{gawedzki1}.

\begin{definition}\label{hbgwc}A hermitian $U(1)$ bundle gerbe $\gerbe$
with connection and curving over a smooth manifold $\targetspace$ consists of the following data: a surjective
submersion $\pi: Y \to \targetspace$, a hermitian
line
bundle $p: L \to Y^{[2] }$ with connection, an associative isomorphism 
\begin{equation}
\mu \maps \pi_{12}^*L\otimes \pi_{23}^*L \lto \pi_{13}^*L
\end{equation}
of hermitian line bundles with connection over $Y^{[3]}$, and a 2-form $C \in \Omega^{2}(Y)$ which satisfies
\begin{equation}
 \pi_{2}^*C-\pi_{1}^*C=\mathrm{curv}(L)\text{.} \end{equation}
\end{definition}
Here $Y^{[p]}$ denotes the $p$-fold fiber
product of $\pi:Y\to\targetspace$, which is
a smooth manifold since $\pi$ is a surjective
submersion.   For example $\pi_{12}:Y^{[3]}\to
Y^{[2]}$ is the projection on the first two
factors. 

\begin{remark}
\label{conventions}From now
we will use the following
conventions: the term line bundle refers to a hermitian line bundle with connection, and  an isomorphism of line bundles
  refers to an isomorphism of hermitian
line bundles
with
connection. Accordingly, we refer to Definition
\ref{hbgwc} by the term
gerbe. The 2-form $C$ 
is called  curving, and  the isomorphism
$\mu$ is called  multiplication. 
\end{remark}

One can show that
there is a unique 3-form $H \in \Omega^{3}(\targetspace)$
with $\pi^*H = \mathrm{d}C$; this 3-form is called
the curvature of the gerbe and is denoted by $H=\mathrm{curv}(\gerbe)$.

To each gerbe $\gerbe$, we associate 
the dual gerbe $\gerbe^{*}$. 
It has the same surjective submersion $\pi : Y \to
\targetspace$, but the dual line bundle $L^{*} \to Y^{´[2]}$ with
multiplication 
\begin{equation}
(\mu^*)^{-1} \maps \pi_{12}^*L^{*}\otimes \pi_{23}^*L^{*} \lto \pi_{13}^*L^{*}\text{,}
\end{equation}
and the negative curving $-C$. Accordingly, the curvature of the dual gerbe
satisfies\begin{equation}
\mathrm{curv}(\gerbe^{*}) = - \mathrm{curv}(\gerbe)\text{.} \end{equation} 
Even more, the classes of $\gerbe$ and the
one of $\gerbe^{*}$ in Deligne hypercohomology
are inverses.

For a smooth map $f:N
\to \targetspace$ and a pullback diagram
\begin{equation}
\label{bg_pb}
\begin{aligned}
\xymatrix{^{}Y_f \ar[d]_{\pi_f} \ar[r]^{\tilde f} & Y \ar[d]^\pi \\ N \ar[r]_f
& \targetspace}
\end{aligned}\text{,}
\end{equation}
  $\pi_f : Y_f \to N$ is a surjective
submersion, and together with the  line bundle
  $\tilde f^{^*}L$ over $Y_f^{[2]}$, the multiplication
$\tilde f^{*}\mu$ and the curving $\tilde f^{*}C$, we have defined
a gerbe $f^{*}\gerbe$. If $f:\targetspace
\to \targetspace$ is a diffeomorphism,
 $Y_f$ is canonically isomorphic to $Y$,
such that $\tilde f = \Id_Y$
and $\pi_f = f^{-1} \circ \pi$. The
curvature of the pullback gerbe is
\begin{equation}
\mathrm{curv}(f^{*}\gerbe)=f^*\mathrm{curv}(\gerbe)\text{.}
\end{equation} 

\begin{remark}
As we did in the last paragraph, whenever there is a map
$\tilde f:Y_f \to Y$, we will use
the same letter for the induced map on higher fiber
products.
\end{remark}
 
\begin{definition}
\label{bg_tr_df}
A trivialization $\mathcal{T}=(T,\tau)$ of a gerbe
$\gerbe$ is a  line bundle $T\to Y$, together with an isomorphism
\begin{equation}
\tau \maps L\otimes \pi_{2}^*T \lto \pi_{1}^*T
\end{equation}
of  line bundles   over $Y^{[2]}$,
which is compatible with the isomorphism $\mu$ of the
gerbe. 
\end{definition}

We call a gerbe $\gerbe$ trivial,
if it admits a trivialization. A choice of
a trivialization $\mathcal{T}$ gives the 2-form $C - \mathrm{curv}(T)\in
\Omega^2(Y)$, which descends to a
unique 2-form $\rho \in \Omega^{2}(\targetspace)$
with $\pi^{*}\rho = C - \mathrm{curv}(T)$.
This 2-form satisfies $\mathrm{d}\rho=H$,
so the curvature $H$ of a trivial gerbe
is an exact form.  

If there are two trivializations $\mathcal{T}_1=(T_1,\tau_1)$
and $\mathcal{T}_2=(T_2,\tau_2)$ of the same gerbe
$\gerbe$, one obtains an isomorphism
\begin{equation}
\alpha :=\tau _1^{-1}\otimes \tau _2^{*}
\maps \pi_{1}^*( T_1\otimes
T_2^*) \lto \pi_2^*( T_1\otimes T_2^*) \text{,}
\end{equation}
of  line bundles  over $Y^{[2]}$. From the compatibility condition between the multiplication $\mu$ and both $\tau_1$ and $\tau_2$ the cocycle condition
\begin{equation}
\pi_{23}^*\alpha \circ \pi_{12}^*\alpha =\pi_{13}^*\alpha 
\end{equation}
follows. Such an isomorphism determines a unique descent line 
bundle $N \to \targetspace $ with connection together with an isomorphism
$\nu : \pi^{*}N \to T_1 \otimes T_2^{*}$ \cite{brylinski1}. The two 2-forms $\rho_1$
and $\rho_2$ coming from the two
trivializations are related by
\begin{equation}
\rho_2 = \rho_1 + \mathrm{curv}(N)\text{.}
\end{equation}

\begin{definition}
\label{bg_si_df}
Let $\gerbe$ and $\gerbe'$ be two gerbes. A stable
isomorphism \begin{equation}\mathcal{A}\maps\gerbe \lto \gerbe'\end{equation}
consists of a  line bundle $A \to Z$ over the fiber product $Z := Y' \times_\targetspace
Y$ 
 with curvature 
\begin{equation}
\mathrm{curv}(A) = p'^*C'-p^*C \text{,}
\label{bg_si_curv}\end{equation}
and an isomorphism
\begin{equation}
\alpha \maps p^*L \otimes p'^*L'^* \otimes \pi_2^*A \lto \pi_1^*A 
\end{equation}
of  line bundles   over $Z^{[2]}$, which is compatible with the multiplications
$\mu$ and $\mu'$ of both gerbes. 
\end{definition}

Here $p$ and $p'$ denote the projections
from $Z$ to $Y$ and to $Y'$ respectively.
Since the pullbacks  of the curvings $C$ and $C'$ to $Z$
differ by a closed 2-form,
the curvatures of stably isomorphic gerbes,  defined by the differential
of $C$, are equal.

\begin{definition}
\label{bg_si_eq}
Let $\gerbe$ and $\gerbe'$ be two
gerbes, and $\mathcal{A}_1$
and $\mathcal{A}_2$
two stable isomorphisms from $\gerbe$
to $\gerbe'$. A morphism 
\begin{equation}
\beta \maps \mathcal{A}_1 \Lto \mathcal{A}_2
\end{equation}
is an isomorphism $\beta:A_1 \to A_2$
of  line bundles over $Z$, which is compatible
with $\alpha_1$ and $\alpha_2$
in the sense that the diagram
\begin{equation}
\begin{aligned}
\xymatrix{
p^{*}L \otimes p'^{*}L'^{*}
\otimes \pi_2^{*}A_1 \ar[rrr]^>>>>>>>>>>>>>>{\alpha_1}
\ar[d]_{1 \otimes 1 \otimes \pi_2^{*}\beta}
&&& \pi_1^{*}A_1 \ar[d]^{\pi_1^{*}\beta} \\ p^{*}L \otimes p'^{*}L'^{*}
\otimes \pi_2^{*}A_2 \ar[rrr]^>>>>>>>>>>>>>>{\alpha_2}
&&& \pi_1^{*}A_2  
}
\end{aligned}
\end{equation}
of isomorphisms of  line
bundles over $Z^{[2]}$ commutes.
\end{definition}
 
The definition of such a morphism of stable
isomorphisms already appeared in \cite{stevenson1}.
We call two stable isomorphisms equivalent,
if there is a morphism between them. This defines an equivalence relation on the
set of stable isomorphisms between two fixed
gerbes $\gerbe$ and $\gerbe'$.

\subsection{Jandl Structures}
\label{ss_js}

Recall that for a group $\orbifoldGroup$
acting on a manifold $\targetspace$ by diffeomorphisms
$\orbifoldGroupElement:\targetspace \to \targetspace$, a $\orbifoldGroup$-equivariant
structure on a  line bundle $L \to \targetspace$
  is a family $\left \lbrace
\varphi^\orbifoldGroupElement
\right \rbrace_{\orbifoldGroupElement \in \orbifoldGroup}$ of isomorphisms
\begin{equation}
\varphi^\orbifoldGroupElement \maps \orbifoldGroupElement^{*}L
\lto L
\end{equation}
of  line bundles,
which respect the group structure of $\orbifoldGroup$
in the sense that $\varphi^{1}:L\to L$ is the identity, and the multiplication law
\begin{equation}
\label{bg_js_es}
\varphi^{\orbifoldGroupElement_1 \orbifoldGroupElement_2}=\varphi^{\orbifoldGroupElement_2}
\circ \orbifoldGroupElement_2^{*}\varphi^{\orbifoldGroupElement_1}
\end{equation}
is satisfied. Remember that according to
our convention in
Remark \ref{conventions}  all line bundles
have connections, and all isomorphisms of
line bundles preserve them. In this article, we  only
consider the group $\orbifoldGroup = \Z_2$
for the sake of simplicity.  
 \vspace{3mm}

Let $\gerbe$ be a gerbe over $\targetspace$
and let $K=\Z_2$ act on $\targetspace$. Denote
the action of  the non-trivial element $\orbifoldGroupElement$ by $\orbifoldGroupElement:\targetspace
\to \targetspace$. Assume that there is a stable isomorphism
$\mathcal{A}=(A,\alpha): \orbifoldGroupElement^{*}\gerbe
\to \gerbe^{*}$. Recall that in this particular
situation, $A$ is a  line bundle
  over the space $Z=Y_\orbifoldGroupElement
\times_\targetspace Y$, where $Y_\orbifoldGroupElement := Y$ and
$\pi_\orbifoldGroupElement := \orbifoldGroupElement^{-1}
\circ \pi$ as in our discussion of the pullback
of $\gerbe$ by a diffeomorphism $\orbifoldGroupElement$.
We still denote the projections from $Z$
to $Y$ and to $Y_\orbifoldGroupElement$ by
$p$ and $p'$ respectively.

Define the surjective submersion $\pi_Z := \pi \circ p :Z \to \targetspace$. As $\orbifoldGroupElement^2=\Id_\targetspace$,
the permutation map \begin{equation}\tilde \orbifoldGroupElement
\maps Z \lto Z \maps (y_\orbifoldGroupElement,y) \mapslto (y,y_\orbifoldGroupElement)\end{equation}
 gives the following commuting diagram:
\begin{equation}
\begin{aligned}
\xymatrix{Z \ar[r]^{\tilde \orbifoldGroupElement} \ar[d]_{\pi_Z} & Z \ar[d]^{\pi_Z} \\ \targetspace \ar[r]_{\orbifoldGroupElement}
& \targetspace}
\end{aligned}
\label{bg_js_li}
\end{equation}
Furthermore, since also $\tilde \orbifoldGroupElement^2=\Id_Z$,
 we even have a lift of the action of $\orbifoldGroup$
 into
$Z$.

\begin{definition}
\label{defJandl}
A Jandl structure on $\gerbe$ is a collection $\mathcal{J}=(\orbifoldGroupElement,\mathcal{A},\varphi)$
consisting of\begin{itemize}
\item 
a smooth action of
$\orbifoldGroup=\Z_2$ on $\targetspace$, where
we denote the non-trivial element
and the  diffeomorphism
associated to that non-trivial element by $\orbifoldGroupElement: \targetspace
\to \targetspace$.
\item a stable isomorphism of gerbes $\mathcal{A}=(A,\alpha):\orbifoldGroupElement^{*}\gerbe \to
\gerbe^{*}$.

\item a $\orbifoldGroup$-equivariant
structure $\varphi:=\varphi^{k}$ on the line
bundle $A$, which
is compatible with the stable isomorphism
$\mathcal{A}$  in the sense that the diagram
\begin{equation}
\begin{aligned}
\xymatrix{p'^*L \otimes p^*L \otimes \pi_2^*A  \ar[rrr]^>>>>>>>>>>>>>>>>{\alpha} \ar[d]_{1 \otimes 1 \otimes \pi_2^*\varphi} &&& \pi_1^*A \ar[d]^{\pi_1^*\varphi} \\ p'^*L \otimes
p^*L \otimes \orbifoldGroupElement^* \pi_2^*A \ar[rrr]^>>>>>>>>>>>>>>{\orbifoldGroupElement^* \alpha}
&&& \orbifoldGroupElement^*\pi_1^*A
}
\end{aligned}
\label{bg_js_co}\end{equation}
of isomorphisms of  line bundles  
over $Z^{[2]}$ commutes.  
\end{itemize} 
\end{definition}

We can immediately deduce a necessary condition for the existence of a Jandl
structure for a given gerbe $\gerbe$, namely the condition, that the gerbes
$\orbifoldGroupElement^{*}\gerbe$ and $\gerbe^{*}$
are stably isomorphic. Since the curvatures of stably
isomorphic gerbes are equal, this in turn demands\begin{equation}
\orbifoldGroupElement^{*}H = -H
\end{equation}
for the curvature $H=\mathrm{curv}(\gerbe)$  of
$\gerbe$. In particular, there will be gerbes
on manifolds with involution which do not
admit a Jandl structure. 

\begin{definition}
Two Jandl structures $\mathcal{J}$ and $\mathcal{J}'$
on the same gerbe $\gerbe$ are equivalent, if the following
conditions are satisfied: 
\begin{itemize}
\item the actions are the same,
i.e.  $\orbifoldGroupElement$ and $\orbifoldGroupElement'$ are the same diffeomorphisms,
\item there is a morphism $\beta:\mathcal{A}
\Rightarrow \mathcal{A}'$  of stable isomorphisms
in the sense
of Definition \ref{bg_si_eq} such
that
\item 
$\beta:A \to A'$ is even an isomorphism
of $\orbifoldGroup$-equivariant  line bundles
on $Z$.
\end{itemize}
\end{definition} 

Next, we show that Jandl structures behave
well under the pullback of gerbes along a smooth map $f:N \to \targetspace$.
Let $\mathcal{J}=(k,\mathcal{A},\varphi)$
be a Jandl structure on $\gerbe$. Assume,
that there is an action of $\orbifoldGroup=\Z_2$ on $N$
by a diffeomorphism $g$, such that
the diagram
\begin{equation}
\begin{aligned}
\label{actionequi}
\xymatrix{N \ar[r]^f \ar[d]_{g} &
\targetspace \ar[d]^\orbifoldGroupElement
\\ N \ar[r]_f & \targetspace &}
\end{aligned}
\end{equation}
commutes. Consider the pullback of $\gerbe$
by $f$ as discussed before, and define
\begin{equation}
Z_f := (Y_f)_g \times_N Y_f 
\end{equation}
and the permutation map $\tilde g: Z_f \to
Z_f$. Then 
\begin{equation}
\begin{aligned}
\xymatrix{&Z_f \ar@{}[dd]_<<<<<<<<<<<<<<<<<{\pi_{Z_f}}
\ar[dd]|<<<<<<<<<<<<{\makebox(5,5){}} \ar[dl]_{\tilde g} \ar[rr]^{\tilde f} & & Z \ar[dl]_{\tilde \orbifoldGroupElement} \ar[dd]^{\pi_Z}\\Z_f \ar[dd]_{\pi_{Z_f}}
 \ar[rr]^>>>>>>>>>>{\tilde f} & & Z \ar[dd]^<<<<<<<<{\pi_Z}&\\&N \ar@{}[rr]^<<<<<<<<<<f \ar[rr]|<<<<<<<<<<<<<{\makebox(5,5){}} \ar[dl]_{g} &&
\targetspace \ar[dl]^\orbifoldGroupElement
\\ N \ar[rr]_f && \targetspace &}
\end{aligned}
\end{equation}
is a cube with commuting faces. It follows that $f^{*}\mathcal{A}
:= (\tilde f^{*}A, \tilde f^{*}\alpha)$ is
a stable isomorphism from $g^{*}f^{*}\gerbe$
to $f^{*}\gerbe^{*}$. Furthermore, $\tilde f^{*}\varphi$ is
a $\orbifoldGroup$-equivariant structure
on $\tilde f^{*}A$, where $\orbifoldGroup$
acts by $\tilde g$.
In summary, 
\begin{equation}
f^{*}\mathcal{J} := (g, \tilde f^{*}\mathcal{A}, \tilde f^{*}\varphi)
\end{equation}
defines a pullback Jandl structure on $f^{*}\gerbe$.

\subsection{Classification of Jandl Structures}
\label{jandlclass}

If a gerbe $\gerbe$ admits a Jandl structure, it is
natural to ask, how many inequivalent choices  exist. So we are interested in
the set $\mathrm{Jdl}(\gerbe, \orbifoldGroupElement)$
of equivalence classes of Jandl
structures $\mathcal{J}=(\orbifoldGroupElement,-,-)$ with a fixed action of $\orbifoldGroup=\Z_2$
via $\orbifoldGroupElement$. This will be
crucial in the discussion of the unoriented
WZW model in  section \ref{ss_wzw}.   

To approach this task, we first investigate
the set $\mathrm{Hom}(\gerbe,\gerbe')$ of equivalence classes of stable isomorphisms between $\gerbe$
and $\gerbe'$. We start by recalling the following

\begin{lemma}[\cite{carey2}]
\label{bg_cl_op}\
\begin{itemize}
\item[(i)] If $N \to \targetspace$ is a flat  line bundle and $\mathcal{A}=(A,\alpha)$
is a stable isomorphism, then $N.\mathcal{A}:=(A
\otimes \pi_Z^{*}N, \alpha \otimes
1)$ is also a stable isomorphism.

\item[(ii)] If $\mathcal{A}_1=(A_1,\alpha_1)$
and $\mathcal{A}_2=(A_2,\alpha_2)$
are two stable isomorphisms, then
there is a unique  flat line bundle
$N \to \targetspace$ such that
$\mathcal{A}_1$ and $N.\mathcal{A}_2$
are equivalent as stable isomorphisms.
\end{itemize}
\end{lemma}

\proof
For the first part we  note that
because $N$ is flat, $A$ and $A
\otimes \pi_{Z}^{*}N$ have the same
curvature, so that (\ref{bg_si_curv})
is satisfied. For the second part,
we use the   isomorphism
\begin{equation}
\alpha_1^{-1} \otimes \alpha_2^* \maps \pi_1^{*}(A_1 \otimes A_2^{*})
\lto \pi_2^{*}(A_1 \otimes A_2^*)
\label{descentiso}\end{equation}
which satisfies the cocycle condition because of
the compatibility of $\alpha_1$ and $\alpha_2$ with
$\mu$ and $\mu'$. This determines a  unique line 
bundle $N \to \targetspace$ with connection together with an isomorphism $\nu: \pi_{Z}^*N \to A_1 \otimes A_2^*$.
Because (\ref{bg_si_curv}) requires the curvatures
of  both 
$A_1$ and $A_2$ to be the same, $N$ is  flat. Now $\nu$
determines an isomorphism $A_1
\to A_2 \otimes \pi_Z^{*}N$, which
is a morphism $\mathcal{A}_1 \Rightarrow
N.\mathcal{A}_2$. 
\endofproof

We denote the group of isomorphism classes
of flat line bundles over $M$ by $\mathrm{Pic}_0(M)$.
It is a subgroup of the Picard group $\mathrm{Pic}(M)$
of isomorphism classes of hermitian  line bundles with connection over
$M$.

\begin{lemma} 
\label{bg_cl_t1}
The set $\mathrm{Hom}(\gerbe,\gerbe')$
of equivalence classes of stable
isomorphisms is a torsor over the
flat Picard group $\mathrm{Pic}_0(M)$.
\end{lemma}

\proof
We will (a) define the action and
show, that it is (b) transitive
and (c) free.
\begin{itemize}
\item[(a)] We act $[N].[\mathcal{A}]:=[N.\mathcal{A}]$,
where the right hand side  was
defined in Lemma \ref{bg_cl_op} (i). This definition is independent
of the choice of representatives
$N$ and $\mathcal{A}$: an isomorphism
$N \to N'$ gives an isomorphism
$N.A \to N'.A$, which in fact is
a morphism of stable isomorphisms
$N.\mathcal{A} \Rightarrow N'.\mathcal{A}$.
On the other hand, a morphism
$\mathcal{A} \Rightarrow \mathcal{A'}$
of stable isomorphisms induces
a morphism $N.\mathcal{A} \Rightarrow
N.\mathcal{A'}$.

Because $N.\mathcal{A}$ is defined
using the group structure on the
group of isomorphism classes of
line bundles with connection, it
respects the group structure
on $\mathrm{Pic}_0(\targetspace)$,
and hence defines an action. 
 
\item[(b)] The transitivity follows
directly from Lemma \ref{bg_cl_op}
(ii).

\item[(c)] Let $[\mathcal{A}]$
 be an
element in $\mathrm{Hom}(\gerbe,\gerbe')$,
let $N$ be a flat 
line bundle and let us assume that
$N.\mathcal{A}$ and $\mathcal{A}$
are equivalent, in particular $A
\otimes \pi_Z^{*}N$ is isomorphic to $A$. Since $N$ is unique
by Lemma \ref{bg_cl_op}
(ii), it is the trivial line bundle. Hence the action
is free.
\endofproof 
\end{itemize}

This lemma allows us to make use of  the flat Picard
group $\mathrm{Pic}_0(\targetspace)$.
Remember that line bundles are, according
to our convention in Remark \ref{conventions}, line bundles with connection. It is well understood \cite{brylinski1},
that the Picard group $\mathrm{Pic}(\targetspace)$
of isomorphism classes of  line bundles 
 fits into the exact
sequence\begin{equation}
\xymatrix{0 \ar[r] & \mathrm{H}^1(\targetspace,U(1))
\ar[r] & \mathrm{Pic}(\targetspace)
\ar[r]^{\mathrm{curv}} & \Omega^2(\targetspace)
}\text{.}
\label{bg_cl_s1}\end{equation}
In particular this means $\mathrm{Pic}_0(\targetspace)
\cong \mathrm{H}^1(\targetspace,U(1))$.
This cohomology group can be computed using
the universal coefficient theorem
\begin{equation}
\xymatrix@=15pt{0 \ar[r] & \mathrm{Ext}(\mathrm{H}_0(M),U(1)) \ar[r] & \mathrm{H}^1(\targetspace,U(1)) \ar[r] & \mathrm{Hom}(\mathrm{H}_1(M),U(1)) \ar[r] &  0 }
\end{equation}
If $M$ is connected, the $\mathrm{Ext}$-group is trivial and
we obtain
\begin{equation}
\mathrm{Pic}_0(\targetspace)\cong\mathrm{Hom}(\pi_1(\targetspace),
U(1))\text{.}
\end{equation}
\vspace{3mm}

An equivariant version of Lemma \ref{bg_cl_t1} applies
to Jandl structures. We denote the group
of isomorphism classes of flat $\orbifoldGroup$-equivariant
line bundles by $\mathrm{Pic}_0^{\orbifoldGroup}(M)$
and call it the flat $\orbifoldGroup$-equivariant
Picard group. In this equivalence relation isomorphisms are isomorphisms of equivariant
line bundles with connection.

\begin{theorem}
\label{bg_js_to}
The set $\mathrm{Jdl}(\gerbe,\orbifoldGroupElement)$
of equivalence classes of Jandl structures
on $\gerbe$ with involution $\orbifoldGroupElement$
is a torsor over the flat $\orbifoldGroup$-equivariant Picard
group $\mathrm{Pic}_0^\orbifoldGroup
(M)$.
\end{theorem}

\proof
\begin{itemize}
\item[(a)] We first describe the action of
a flat  line bundle $N$ over $\targetspace$
with equivariant structure $\nu$ on a Jandl structure $\mathcal{J}=(k,\mathcal{A},\varphi)$.
According to diagram
(\ref{bg_js_li}), $\pi_Z^{*}\nu: \pi_Z^{*}N
\to \tilde \orbifoldGroupElement^{*}\pi_Z^{*}N$ is a $\orbifoldGroup$-equivariant
structure on $\pi_Z^{*}N$.
Now, by taking the tensor product of $A$
and $\pi_Z^{*}N$ as $\orbifoldGroup$-equivariant line bundles, we obtain an equivariant structure
$\varphi \otimes \pi_Z^{*}\nu$ on the line
bundle of $N.\mathcal{A}$. So we define
\begin{equation}
N.\mathcal{J} := (\orbifoldGroupElement,
N.\mathcal{A}, \varphi \otimes \pi_Z^{*}\nu)\text{.}
\end{equation}
Since
\begin{equation}
\begin{aligned}
\xymatrix{Z^{[2]} \ar[r]^{\pi_1} \ar[d]_{\pi_2}
& Z \ar[d]^{\pi_{Z}} \\ Z \ar[r]_{\pi_{Z}} & M}
\end{aligned}
\end{equation}
commutes, we have $\pi_1^{*}\pi_Z^{*}\nu=\pi_2^{*}\pi_Z^{*}\nu$.
This shows that condition (\ref{bg_js_co})
for Jandl structures is satisfied for $N.\mathcal{J}$.
The arguments in the proof of Lemma \ref{bg_cl_t1}
(a)
apply here too and show that this
defines an action on equivalence classes. 
\item[(b)] Let two equivalence
classes of Jandl structures be
represented by $\mathcal{J}_1$ and
$\mathcal{J}_2$. We already know
from Lemma \ref{bg_cl_op} (ii)
that there is a  flat line
bundle $N \to \targetspace$ together with
an isomorphism $\beta :A_1
\to A_2 \otimes \pi_Z^{*}N$, which is a morphism
of stable isomorphism $\beta :N.\mathcal{A}_1 \Rightarrow \mathcal{A}_2$. We have to show
that  there is an equivariant structure on
$N$ such that $\beta$ is an isomorphism of
equivariant line bundles. Remember that we
defined $N$ by a descent isomorphism $\alpha_1^{-1} \otimes \alpha_2^*$ in (\ref{descentiso}).
Because the equivariant structures on $A_{1}$
and $A_{2}$ are compatible with $\alpha_{1}$
and $\alpha_{2}$ respectively due to the
property (\ref{bg_js_co}) of Jandl structures,
the descent isomorphism is an isomorphism
of equivariant line bundles. Thus $N$ is
an equivariant line bundle, and $\beta$ is
an isomorphism of equivariant line bundles.

\item[(c)] Let $\jandl =(\orbifoldGroupElement,
\mathcal{A}, \varphi)$ represent a  Jandl
structure on $\gerbe$, and let $N$ be a flat
line bundle over $\targetspace$
with equivariant structure $\nu$,
such that $N.\jandl$ and $\jandl$
 are equivalent. It follows  from
 Lemma \ref{bg_cl_t1} that $N$
 is the trivial line bundle. Furthermore,
 $\pi_Z^{*}\nu$ is the trivial
 equivariant structure on $\pi_Z^{*}N$,
 so that $\nu$ is the trivial equivariant structure on $N$. \endofproof \end{itemize}

For an action
of a discrete group $\orbifoldGroup$ on $\targetspace$,
an equivariant version of the sequence
(\ref{bg_cl_s1}) is derived in \cite{gomi1}, namely
\begin{equation}
\xymatrix{0 \ar[r] & \mathrm{H}_\orbifoldGroup^1(\targetspace,U(1))
\ar[r] & \mathrm{Pic}^\orbifoldGroup(\targetspace)
\ar[r]^{\mathrm{curv}} & \Omega^2(\targetspace)^\orbifoldGroup
}\text{.}
\end{equation}
Here, $\mathrm{H}_\orbifoldGroup^1(\targetspace,U(1))$
is the equivariant cohomology of
$\targetspace$, i.e. the cohomology
of the associated Borel space.
In particular, we get for flat
equivariant line bundles 
\begin{equation}
\label{bg_js_pic}
\mathrm{Pic}_0^\orbifoldGroup(\targetspace)\cong\mathrm{H}_\orbifoldGroup^1(\targetspace,U(1))\text{.}
\end{equation}

\subsection{Local Data}
\label{ss_ld}

Let $\gerbe$ be a gerbe over $\targetspace$
and $\mathfrak{\targetspaceCover}=\left \lbrace \targetspaceCover_i \right \rbrace_{i \in I}$ be a
good open cover of $\targetspace$.
Let $\targetspaceNerve$
be the disjoint union of all the
$\targetspaceCover_i$'s. The $p$-fold
fiber product of $\targetspaceNerve$
over $\targetspace$ is just the
disjoint union of all $p$-fold
intersections of the $\targetspaceCover_i$'s. Recall from \cite{carey2} how to
extract local data from $\gerbe$:

A choice of local sections $s_i:\targetspaceCover_i
\to Y$ gives a fiber preserving
map $s:\targetspaceNerve
\to Y$ by $(x,i) \mapsto s_i(x)$.
 Pull back the line bundle $L \to
 Y^{[2]}$ with its connection $\nabla$
 along $s$ to a line bundle on the double
 intersections, and
 choose local sections $\sigma_{ij}:
 \targetspaceCover_i \cap \targetspaceCover_j \to s^{*}L$. Pull back the isomorphism
$\mu$ of the gerbe, too. Then define
local data, namely smooth
functions $\gerbelocijk_{ijk}:\targetspaceCover_i
\cap \targetspaceCover_j \cap \targetspaceCover_k
\to U(1)$, real-valued 1-forms $\gerbelocij_{ij}
\in \Omega^1(\targetspaceCover_i
\cap \targetspaceCover_j)$ and
2-forms $\gerbeloci_i \in \Omega^2(\targetspaceCover_i)$
by the following relations
\begin{eqnarray}
s^{*}\mu\left( \pi_{12}^* \sigma _{ij}\otimes \pi_{23}^*\sigma
_{jk}\right) &=&\gerbelocijk_{ijk}\cdot \pi_{13}^*
\sigma _{ik}  \\
s^{*}\nabla(\sigma _{ij}) &=&\frac{1}{\im}\gerbelocij_{ij}\otimes \sigma _{ij} \\
\gerbeloci_{i} &=&s_{i}^* C \text{.} \end{eqnarray}
These local data give elements
$\gerbelocijk$, $\gerbelocij$, $\gerbeloci$ in the \v Cech-Deligne
double complex for the cover $\mathfrak{\targetspaceCover}$, and the cochain
$(\gerbelocijk, \gerbelocij, \gerbeloci)$ satisfies the Deligne
cocycle condition 
\begin{equation}
\deligneDifferential\left( \gerbelocijk,\gerbelocij,\gerbeloci\right) =\left( 1,0,0\right)\text{,}
\end{equation}
or equivalently in components
\begin{eqnarray}
\gerbelocijk_{jkl}\cdot \gerbelocijk_{ikl}^{-1}\cdot \gerbelocijk_{ijl}\cdot \gerbelocijk_{ijk}^{-1} &=&1
\\
\gerbelocij_{jk}-\gerbelocij_{ik}+\gerbelocij_{ij}+\dlog\left( \gerbelocijk_{ijk}\right) &=&0
\\
-\exteriorDifferential \gerbelocij_{ij}+\gerbeloci_{j}-\gerbeloci_{i} &=&0\text{.}   
\end{eqnarray}
Furthermore, it satisfies \begin{equation}
\exteriorDifferential\gerbeloci_{i} = H|_{\targetspaceCover_i}\text{,}
\end{equation}
where the 3-form $H$ is the curvature
of the gerbe.

The dual gerbe and the pullback gerbe $f^{*}\gerbe$ along some map $f:N \to \targetspace$ can
be conveniently expressed in local data as
follows: by choosing the same $s_i$
and the dual sections $\sigma^{*}_{ij}$, one gets $(\gerbelocijk^{-1}, -\gerbelocij, -\gerbeloci)=-(\gerbelocijk, \gerbelocij, \gerbeloci)$ as local data of $\gerbe^{*}$. Furthermore, if we induce a cover $\lbrace
f^{-1}\targetspaceCover_i \rbrace_{i \in I}$ of $N$, and choose the pullback sections $f^{*}s_i$ and
$\tilde f^{*}\sigma_{ij}$, then we obtain $(f^{*}\gerbelocijk, f^{*}\gerbelocij, f^{*}\gerbeloci)=f^{*}(\gerbelocijk, \gerbelocij, \gerbeloci)$ as local data of $f^{*}\gerbe$.

\vspace{3mm}

We next need to derive local data of trivializations
and stable isomorphisms. So, let
$\mathcal{T}=(T,\tau)$ be a trivialization
of $\gerbe$. Since $T$ is a line
bundle over $Y$, we can pull it
back with $s: \targetspaceNerve
\to Y$ to a line bundle over the open subsets, and choose local sections
$\sigma_{i}: \targetspaceCover_i \to
s^{*}T$. We also pull back the
isomorphism $\tau$ to an isomorphism
\begin{equation}
s^{*}\tau \maps s^{*}L \otimes \pi_{2}^*s^{*}T \lto \pi_{1}^*s^{*}T\text{.}
\end{equation}  
Then we obtain smooth functions
$\gerbetrivij_{ij}:\targetspaceCover_i
\cap \targetspaceCover_j \to U(1)$
by
\begin{equation}
s^{*}\tau \left( \sigma _{ij} \otimes
\pi_{2}^* \sigma _{j}\right)
=\gerbetrivij_{ij}\cdot \pi_{1}^{*}\sigma
_{i}\text{.}
\end{equation}
Let $\blacktriangledown$ be the
connection of $T$. It defines connection
1-forms
$\gerbetrivi_{i} \in \Omega^1(\targetspaceCover_i
)$ by
\begin{equation}
s^{*}\blacktriangledown \left( \sigma _{i}\right) =\frac{1}{\im}
\gerbetrivi_{i}\otimes
\sigma _{i}\text{.}
\end{equation}
The local data $\gerbetrivij$ and
$\gerbetrivi$ are again elements
in the \v Cech-Deligne double complex.
Now the compatibility of $\tau$
and $\mu$ in Definition \ref{bg_tr_df}
is equivalent to
\begin{equation}
\gerbelocijk_{ijk}=\gerbetrivij_{ij}\cdot \gerbetrivij_{ik}^{-1}\cdot \gerbetrivij_{jk}\text{,}
\end{equation}
and the condition, that the isomorphism $\tau$ respect connections, is
equivalent to
\begin{equation}
\gerbelocij_{ij}=-\dlog\left( \gerbetrivij_{ij}
\right) +\gerbetrivi_{j}-\gerbetrivi_{i}
\text{.}\end{equation}
Furthermore, the local 2-form $\rho
= \gerbeloci_i + \exteriorDifferential
\gerbetrivi_i$ coincides with the
2-form $\rho$ obtained from Definition
\ref{bg_tr_df}. The last three
properties of $\gerbetrivij$ and $\gerbetrivi$ are equivalent to
the Deligne coboundary equation
\begin{equation}
\left(\gerbelocijk,\gerbelocij,\gerbeloci
\right) =\left( 1,0,\rho \right) +\deligneDifferential\left( \gerbetrivij,
\gerbetrivi
\right)
\text{.}
\end{equation}

Now consider a stable isomorphism $\mathcal{A}:\gerbe \to \gerbe'$ of gerbes over $\targetspace$. With
respect to the good open cover $\lbrace \targetspaceCover_i \rbrace_{i \in I}$ we may have chosen local
sections $s_i$, $\sigma_{ij}$ and $s'_i$, $\sigma'_{ij}$ to get local data $(\gerbelocijk,
\gerbelocij, \gerbeloci)$ and $(\gerbelocijk',
\gerbelocij', \gerbeloci')$ of $\gerbe$ and $\gerbe'$ respectively. We construct a map
\begin{equation}
\label{bg_ld_st}
\tilde s \maps \targetspaceNerve \lto Y \times_\targetspace Y' \maps (x,i) \mapslto (s_i(x) , s'_i(x)) \text{,}
\end{equation} 
and pull the line bundle $A \to Y \times_M Y'$ of the stable isomorphism together with its connection
$\blacktriangledown$ back to $\targetspaceNerve$.
We also pull back the isomorphism $\alpha$ and get an isomorphism
\begin{equation}
\tilde s^*\alpha \maps s^{*}L \otimes s'^*L'^{*} \otimes \pi_2^{*}\tilde s^{*}A \lto \pi_1^{*} \tilde s^{*}A\text{.}
\end{equation}
Then we choose local sections  $\sigma_i : \targetspaceCover_i \to \tilde s^*A$. We obtain local data
in form of smooth functions
$\jandllocij_{ij}: \targetspaceCover_i \cap \targetspaceCover_j \to U(1)$ and connection
1-forms $\jandlloci_i
\in \Omega^1(\targetspaceCover_i)$ by the following relations:
\begin{eqnarray}
\tilde s^*\alpha \left(\sigma _{ij}\otimes \sigma_{ij}^{\prime *}\otimes \pi_2^*\sigma _{j}^{*}\right) &=&\jandllocij_{ij}\cdot \pi_1^*\sigma _{i} \\
\tilde s^* \blacktriangledown ( \sigma _{i}) &=&\frac{1}{\im} \jandlloci_{i}\otimes \sigma _{i}\text{.} 
\end{eqnarray}
Note that the functions $\jandllocij_{ij}$
are not transition functions of some bundle
but  are defined by the isomorphism $\alpha$.

 These local data $\jandllocij$ and $\jandlloci$ are elements in the \v Cech-Deligne double complex.
The compatibility of $\alpha$ with the isomorphisms $\mu$ and $\mu'$ of both gerbes as isomorphisms
of hermitian line bundles with connection according to Definition
\ref{bg_si_df} is equivalent to 
\begin{eqnarray}
\gerbelocijk_{ijk}\cdot \gerbelocijk_{ijk}^{\prime -1} &=&\jandllocij_{jk} \cdot \jandllocij_{ik}^{-1} \cdot \jandllocij_{ij} \\
\gerbelocij_{ij}-\gerbelocij'_{ij}&=&-\dlog (\jandllocij_{ij}) + \jandlloci_{j}- \jandlloci_{i}
\end{eqnarray}
while the condition (\ref{bg_si_curv}) on the curvature of $A$ is equivalent to
\begin{equation}
\gerbeloci_{i}-\gerbeloci'_{i}=\exteriorDifferential \jandlloci_{i} \text{.}
\end{equation}
The three last equations are in turn equivalent to the Deligne coboundary equation
\begin{equation}
\label{bg_ld_si}
\left(\gerbelocijk, \gerbelocij, \gerbeloci \right) -\left( \gerbelocijk', \gerbelocij', \gerbeloci'
\right) = \deligneDifferential \left(\jandllocij,\jandlloci\right)\text{.}
\end{equation}

This formalism of local data reproduces results on bundle gerbes and their stable isomorphisms,
for example Lemma \ref{bg_cl_op} (ii). Consider again two gerbes $\gerbe$ and $\gerbe'$, and now two
stable isomorphisms $\mathcal{A}_1$ and $\mathcal{A}_2$ both from $\gerbe$ to $\gerbe'$. We may have
extracted local data $(\jandllocij_1, \jandlloci_1)$ of $\mathcal{A}_1$ and $(\jandllocij_2, \jandlloci_2)$ of $\mathcal{A}_2$ such that equation (\ref{bg_ld_si}) holds for both. It follows
\begin{equation}
\deligneDifferential(\jandllocij \cdot \jandllocij'^{-1}, \jandlloci - \jandlloci') = (1, 0, 0)\text{,}
\end{equation}
which is the Deligne cocycle condition for a flat hermitian line bundle over $\targetspace$. This is
the  bundle $N$ constructed in Lemma \ref{bg_cl_op} (ii).

\vspace{3mm}
We are now in a position to derive the local data of a Jandl structure $\mathcal{J}=(\orbifoldGroupElement, \mathcal{A}, \varphi)$ on a gerbe $\gerbe$. Recall that $\orbifoldGroupElement:\targetspace \to \targetspace$ is the action of the non-trivial element of $\orbifoldGroup=\Z_2$ acting on $\targetspace$, in particular $\orbifoldGroupElement^2=\Id_\targetspace$. We simplify the situation by considering
an open cover $\mathfrak{\targetspaceCover}
= \lbrace \targetspaceCover_i \rbrace_{i \in I}$ of $\targetspace$, which is invariant under $\orbifoldGroupElement$, i.e. $\orbifoldGroupElement(\targetspaceCover_i)
= \targetspaceCover_i$, and which is still good enough to enable us to extract local data. The generalization other covers is straightforward, but makes
the notation somewhat more cumbersome. 

Recall further that $\mathcal{A}$ is a stable isomorphism from $\orbifoldGroupElement^{*}\gerbe \to \gerbe^{*}$. Let $(\jandllocij, \jandlloci)$ be local data of $\mathcal{A}$, obtained by pulling back
the line bundle $A \to Z$ by $\tilde s: \targetspaceNerve \to Z$  from equation (\ref{bg_ld_st}) and
choosing local sections $\sigma_i: \targetspaceCover_i \to \tilde s^{*}A$. As we derived for the
local data of   the dual gerbe and the pullback gerbe, equation (\ref{bg_ld_si}) here appears as 
\begin{equation}
\orbifoldGroupElement^{*} (\gerbelocijk, \gerbelocij, \gerbeloci) = - (\gerbelocijk, \gerbelocij, \gerbeloci)
+ \deligneDifferential(\jandllocij, \jandlloci)\text{,}
\end{equation}
or equivalently:
\begin{eqnarray}
\orbifoldGroupElement^{*}\gerbeloci_{i} &=& - \gerbeloci_{i}
+\exteriorDifferential \jandlloci_{i} \\
\orbifoldGroupElement^{*} \gerbelocij_{ij} &=& - \gerbelocij_{ij}
 -\dlog(\jandllocij_{ij})
+ \jandlloci_{j} - \jandlloci_{i} \\
\orbifoldGroupElement^{*}\gerbelocijk_{ijk} &=& \gerbelocijk_{ijk}^{-1}
\cdot t_{jk} \cdot t_{ik}^{-1} \cdot t_{ij} 
\end{eqnarray}

Now recall that a part of a Jandl structure
is a $\orbifoldGroup$-equivariant structure
 $\varphi: \orbifoldGroupElement^*A \to A$
on $A$. By pullback with $\tilde s$, we obtain
\begin{equation}
\tilde s^{*}\varphi \maps k^{*}\tilde s^{*}A \lto \tilde s^{*}A\text{.}
\end{equation}
Now, because $\sigma_i$ is a section of $\tilde s^{*}A$, $\orbifoldGroupElement^{*}\sigma_i = \sigma_i
\circ \orbifoldGroupElement$ is a section of $\orbifoldGroupElement^{*} \tilde s^{*} A$ on the same
patch $\targetspaceCover_i$, since the latter is invariant under $\orbifoldGroupElement$. This allows us to
extract a local $U(1)$-valued functions $\jandlloccc_i: \targetspaceCover_i \to U(1)$, defined by
\begin{equation}
\tilde s^{*} \varphi(\sigma_i)= \jandlloccc_i \cdot \sigma_i \circ \orbifoldGroupElement\text{.}
\end{equation}
The compatibility of $\varphi$ with $\alpha$ in the sense of diagram (\ref{bg_js_co}) is equivalent
to
\begin{equation}
\orbifoldGroupElement ^{*}\left(\jandllocij , \jandlloci \right) =\left( \jandllocij , \jandlloci \right) -\deligneDifferential \left( \jandlloccc \right) \text{,}
\end{equation}
or in turn equivalently
\begin{eqnarray}
\orbifoldGroupElement ^{*} \jandlloci _{i} &=& \jandlloci _{i} - \dlog (\jandlloccc_{i}) \\
\orbifoldGroupElement ^{*}\jandllocij_{ij}  &=&  \jandllocij_{ij} \cdot \jandlloccc_{j}^{-1}
 \cdot \jandlloccc_{i} \text{.}\end{eqnarray}

By definition of an equivariant structure, the $\orbifoldGroup=\Z_2$ group law (\ref{bg_js_es}) is satisfied. In terms of local data, this is equivalent to
\begin{equation}
\orbifoldGroupElement^{*}\jandlloccc_i = \jandlloccc_i^{-1}\text{.}
\end{equation}

In summary, the Jandl structure $\mathcal{J}=(\orbifoldGroupElement, \mathcal{A}, \varphi)$ gives rise
to local data $(\jandllocij, \jandlloci)$ and $\jandlloccc$ which satisfy the following three conditions:
\begin{eqnarray}
\label{bg_ld_c1}
\orbifoldGroupElement^{*} (\gerbelocijk, \gerbelocij, \gerbeloci) &=& - (\gerbelocijk, \gerbelocij, \gerbeloci)
+ \deligneDifferential(\jandllocij, \jandlloci) \label{bg_ld_1}\\
\orbifoldGroupElement ^{*}\left(\jandllocij , \jandlloci \right) &=& \left( \jandllocij , \jandlloci \right) -\deligneDifferential \left( \jandlloccc \right ) \label{bg_ld_2}\\
\orbifoldGroupElement^{*}\jandlloccc_i &=& \jandlloccc_i^{-1}
\label{bg_ld_3}\end{eqnarray}

\vspace{3mm}

Again, using  local data, we can reproduce results on Jandl structures like Theorem \ref{bg_js_to}.
In detail, let $\mathcal{J}$ be a Jandl structure on $\gerbe$ with local data $(\jandllocij, \jandlloci
)$ and $\jandlloccc$. Let $N$ be a flat $\orbifoldGroup$-equivariant hermitian line bundle
 over $\targetspace$ with transition functions $n_{ij}:
\targetspaceCover_i \cap \targetspaceCover_j \to U(1)$ and local connection 1-forms $N_i \in \Omega^1(\targetspaceCover_i)$
with
\begin{equation}
\deligneDifferential(n,N)=(1,0,0).
\label{bg_ld_cn}\end{equation}
The equivariant structure on $N$ determines smooth functions $\nu_i : \targetspaceCover_i \to U(1)$ with
\begin{equation}
\orbifoldGroupElement^{*}(n,N) = (n,N) - \deligneDifferential(\nu)
\end{equation}
and $\orbifoldGroupElement^{*}\nu = \nu^{-1}$. Then,
\begin{eqnarray}
(\jandllocij', \jandlloci') &:=& (\jandllocij, \jandlloci) + (n,N) \\
\jandlloccc' &:=& \jandlloccc \cdot \nu
\end{eqnarray}
are local data of the Jandl structure $N.\mathcal{J}$. Indeed, equation (\ref{bg_ld_1}) is satisfied
because of the Deligne cocycle condition (\ref{bg_ld_cn}). Compute
\begin{eqnarray}
\nonumber \orbifoldGroupElement^{*}(\jandllocij', \jandlloci') &=& \orbifoldGroupElement^{*}(\jandllocij, \jandlloci)
+ \orbifoldGroupElement^{*} (n,N)\\\nonumber  &=& \left( \jandllocij , \jandlloci \right) -\deligneDifferential \left( \jandlloccc \right ) + (n,N) - \deligneDifferential(\nu) \\ &=& (\jandllocij', \jandlloci') -
\deligneDifferential(\jandlloccc') \text{,}
\end{eqnarray}  
this is equation (\ref{bg_ld_2}), and the last equation (\ref{bg_ld_3}) for $\jandlloccc'$
is just a consequence from the
conditions on $\jandlloccc$ and $\nu$.

Let now $\jandl$ and $\jandl'$ be two Jandl structures on $\gerbe$ with local data $(\jandllocij, \jandlloci),
\jandlloccc$ and $(\jandllocij', \jandlloci'), \jandlloccc'$ respectively. 
\begin{equation}
(n, N) := (\jandllocij, \jandlloci) - (\jandllocij', \jandlloci')
\end{equation}
are the local data of the flat descent line bundle $N$, and using equation (\ref{bg_ld_1}),
we get its cocycle condition 
\begin{equation}
\deligneDifferential(n,N)=(1,0,0)\text{.}
\end{equation}    
Now compute
\begin{eqnarray}
\orbifoldGroupElement^{*}(n,N) &=& \orbifoldGroupElement^{*}(\jandllocij, \jandlloci) - \orbifoldGroupElement^{*}(\jandllocij', \jandlloci')\nonumber  \\ &=& (\jandllocij, \jandlloci) - \deligneDifferential(\jandlloccc) - (\jandllocij', \jandlloci')
+  \deligneDifferential(\jandlloccc')\nonumber  \\ &=& (n,N) - \deligneDifferential(\nu)\text{,}
\end{eqnarray}
where we defined $\nu := \jandlloccc \cdot \jandlloccc'^{-1}$. Hence, $N$ and $\orbifoldGroupElement^{*}N$
are isomorphic as hermitian line bundles with connection via an isomorphism represented by $\nu$. By
definition, we have $\orbifoldGroupElement^{*}\nu = \nu^{-1}$, this means, that $\nu$ is a
$K$-equivariant
structure.

\section{Holonomy of Gerbes
with Jandl Structure}
\label{ss_hol}

\subsection{Double Coverings, Fundamental
Domains and Orientations}
\label{ss_fd}

Let us first recall the setup that allows
to define
holonomy around closed oriented  surfaces.
This is a gerbe $\gerbe$ over $\targetspace$ and a closed oriented
surface $\worldsheet$
together with a smooth map $\stringmap:
\worldsheet \to \targetspace$.
Following \cite{carey2}, we pull back
 $\gerbe$ along $\stringmap$ 
 to a gerbe over $\worldsheet$. For dimensional
 reasons, $\stringmap^{*}\gerbe$ is trivial.
As explained in section \ref{ss_bg_bg},
a trivialization
$\mathcal{T}$ determines a 2-form
$\rho \in \Omega^2(\worldsheet)$,
while another trivialization $\mathcal{T}'$
determines a 2-form $\rho' = \rho
+ \mathrm{curv}(N)$. Since $\mathrm{curv}(N)$
defines an integral class in cohomology,
we have
\begin{equation}
\int_\worldsheet \rho' = \int_\worldsheet \rho \;\; \mathrm{mod} \;\; 2\pi\Z\text{.}
\end{equation}
So the integral is independent
of the choice of a trivialization
up to $2\pi\Z$, and admits therefore
the following
\begin{definition}
\label{hol_bg_or}The holonomy of $\gerbe$ around
the closed oriented surface $\stringmap:
\worldsheet \to \targetspace$ is
defined
as
\begin{equation}
\mathrm{hol}_\gerbe(\stringmap,\worldsheet):=\exp
\left ( \im \int_\worldsheet
\rho \right ) \in U(1)\text{.}
\end{equation}
\end{definition}
\noindent We state three important properties of this
definition:
\begin{itemize}

\item[$\bullet$]
The dual gerbe has inverse holonomy,
\begin{equation}\mathrm{hol}_\gerbe(\stringmap,\worldsheet)
= \mathrm{hol}_{\gerbe^{*}}(\stringmap,\worldsheet)^{-1}\text{.}\end{equation}

\item[$\bullet$]
If $\mathcal{A}:\gerbe
\to \gerbe'$ is a stable isomorphism, we
have 
\begin{equation}
\mathrm{hol}_\gerbe(\stringmap,\worldsheet)=\mathrm{hol}_{\gerbe'}(\stringmap,\worldsheet)\text{.}
\end{equation}

\item[$\bullet$]
By $\bar \worldsheet$ we denote the same manifold
$\worldsheet$ with the opposite orientation;
then we obtain
\begin{equation}
\mathrm{hol}_\gerbe(\stringmap,\worldsheet)=\mathrm{hol}_{\gerbe}(\stringmap,\bar\worldsheet)^{-1}\text{.}
\end{equation}

\end{itemize}

Obviously, the orientation on $\worldsheet$
is essential for this definition.
In this section we will define
the holonomy around unoriented
or even unorientable surfaces.
The most important property of
this definition will be, that it
reduces to Definition \ref{hol_bg_or}
if $\worldsheet$ is orientable and an orientation
is chosen. One of the main tools will be an orientation covering.

Let $\worldsheet$ be a smooth manifold (without
orientation). 

\begin{definition}
An orientation covering  
of $\worldsheet$ is a double covering $\worldsheetCovering:
\worldsheetCoveringSpace \to \worldsheet$ with an oriented manifold $\worldsheetCoveringSpace$,
such that the canonical
involution $\worldsheetInvolution:
\worldsheetCoveringSpace \to \worldsheetCoveringSpace$
is orientation-reversing.
\end{definition}

Recall three  basic properties of  orientation
coverings (some of them can be found for example in \cite{berger1}):
\begin{itemize}
\item it
is unique up to orientation-preserving diffeomorphisms
of covering spaces.

\item the canonical involution
$\worldsheetInvolution:
\worldsheetCoveringSpace \to \worldsheetCoveringSpace$
 preserves fibers and permutes
the the sheets.

\item under the assumption that
$\worldsheet$ is connected, $\worldsheetCoveringSpace$
is connected if and only if $\worldsheet$
is not orientable. 
\end{itemize}

Due to the first point, by $\worldsheetCoveringSpace$ we will from now refer to this unique orientation
cover. Let $\orbifoldGroupElement:
\targetspace \to \targetspace$ be an involution
on $\targetspace$.
By $C^{\infty}(\worldsheetCoveringSpace,M)^{\worldsheetInvolution,k}$
we denote the space of smooth maps $\hat
\stringmap:
\worldsheetCoveringSpace \to \targetspace$
for which the diagram 
\begin{equation}
\label{bg_hol_co}
\begin{aligned}
\xymatrix{\worldsheetCoveringSpace
\ar[r]^{\hat \stringmap} \ar[d]_{\worldsheetInvolution} & \targetspace \ar[d]^{k}
\\ \worldsheetCoveringSpace
\ar[r]_{\hat \stringmap} & \targetspace}
\end{aligned}
\end{equation} 
commutes in the category of smooth manifolds
(neglecting orientations).

Let $\worldsheet$ be orientable. 

\begin{lemma}
\label{hol_bg_bi}
An orientation on $\worldsheet$
defines a bijection
\begin{equation}
C^{\infty}(\worldsheetCoveringSpace,M)^{\worldsheetInvolution,k}
\lto C^{\infty}(\worldsheet,M).
\end{equation} 
\end{lemma}

\proof
Since $\worldsheet$ is orientable,
$\worldsheetCoveringSpace$ consists
of two disjoint copies of $\worldsheet$
with opposite orientations. An
orientation on $\worldsheet$ is a global
section
$\mathrm{or}:\worldsheet \to \worldsheetCoveringSpace$
in the covering $\worldsheetCovering:\worldsheetCoveringSpace
\to \worldsheet$. 
Now let $\hat \stringmap:\worldsheetCoveringSpace
\to \targetspace$ be a map. Define its image
as $\stringmap
:= \hat  \stringmap \circ \mathrm{or}$. On the other hand, given
a map $\stringmap: \worldsheet \to \targetspace$,
we define the preimage $\hat \stringmap$ on the two sheets of $\worldsheetCoveringSpace$
separately as
$\hat \stringmap|_{\mathrm{or}(\worldsheet)}:=\stringmap$
and $\hat \stringmap|_{\worldsheetInvolution
\mathrm{or}(\worldsheet)}:= k \circ \stringmap$
respectively.   
\endofproof

If $\worldsheet$ is not orientable or no
orientation of $\worldsheet$ is chosen, we will
make
use of the following generalization of an
orientation.

\begin{definition}
A fundamental domain for   $\worldsheet$ in  
$\worldsheetCoveringSpace$ is
a submanifold $F \subset \worldsheetCoveringSpace$
possibly with (piecewise smooth) boundary, satisfying the following
two conditions as sets:
\begin{itemize}
\item[(i)] $F \cap \worldsheetInvolution(F)
= \partial F$
\item[(ii)] $F \cup \worldsheetInvolution(F)=\worldsheetCoveringSpace$
\end{itemize}
\end{definition}

This is a generalization of an orientation
on $\worldsheet$ in the sense, that any orientation
on $\worldsheet$ gives a global section
$\mathrm{or}: \worldsheet \to \worldsheetCoveringSpace$
which in turn defines a fundamental domain, namely $F:=\mathrm{or}(\worldsheet)$,
one of the two copies of $\worldsheet$
in $\worldsheetCoveringSpace$.

We  show the existence of such a fundamental
domain for an arbitrary closed surface $\worldsheet$
by an explicit construction, which we will
also use in section \ref{ss_hol_ld}. Let
$\mathfrak{\worldsheetCover}=\lbrace \worldsheetCover_i \rbrace_{i \in
I}$ be an open cover of $\worldsheet$, which
admits local sections $\mathrm{or}_i:\worldsheetCover_i
\to \worldsheetCoveringSpace$. One can think
of such sections as local orientations. Choose a dual triangulation $T$ of $\worldsheet$, subordinate
to  the cover $\mathfrak{\worldsheetCover}$, together with a subordinating map $i:
T \to I$. So, for each face $f \in T$
there is an index $i(f)$ with $f
\subset \worldsheetCover_{i(f)}$, as
well as for each edge $e \in T$ and for each
vertex $v \in T$. Because we have a dual
triangulation, each vertex is trivalent.

Consider a common edge $e = f_1 \cap f_2$
of two faces $f_1$ and $f_2$. We call the
edge  $e$
orientation-preserving, if 
\begin{equation}
\mathrm{or}_{i(f_1)}(e)=\mathrm{or}_{i(f_2)}(e)\text{,}
\end{equation}
otherwise we call it orientation-reversing.
So the set of edges splits in a set $E$
of orientation-preserving, and a
set $\bar E$ of orientation-reversing edges.
If $v$ is a vertex, the number of orientation-reversing
edges ending in $v$ must be even, and since
we started with a dual triangulation, it
is either zero or two. Hence, the edges in $\bar E$ form non-intersecting closed lines in $\worldsheet$.

\begin{figure}[h]
\begin{center}
\includegraphics{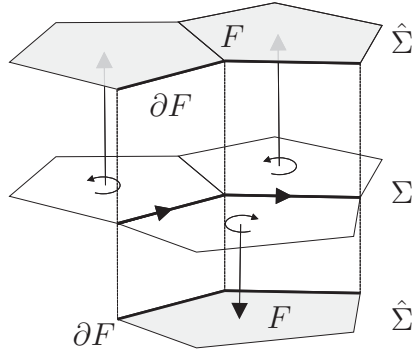}\setlength{\unitlength}{1pt}\begin{picture}(0,0)(204,392)\put(116.68394,506.75074){$\worldsheetCoveringSpace$}\put(70.24394,402.76885){$F$}\put(52.66346,508.85688){$F$}\put(116.63887,449.94813){$\worldsheet$}\put(116.63887,402.30992){$\worldsheetCoveringSpace$}\put(-2.38139,396.44135){$\partial F$}\put(26.41861,482.84135){$\partial F$}\end{picture}
\caption{The construction of a
fundamental domain by local orientations
for a dual triangulation. }
\end{center}
\end{figure}

Define the subset
\begin{equation}
F := \bigcup_{f \in T} \mathrm{or}_{i(f)}(f).
\end{equation}
of $\worldsheetCoveringSpace$ and endow it
with the subspace topology. The boundary
of $F$ is exactly the union of the preimages of orientation-reversing edges under the
covering map,
\begin{equation}
\partial F = \bigcup_{e \in \bar E} \worldsheetCovering^{-1}(e)\text{,}
\end{equation}
and hence  a disjoint union of piecewise smooth circles. This shows that $F$ is a submanifold  of $\worldsheetCoveringSpace$
with piecewise smooth boundary. It satisfies the
two properties of a fundamental domain, and hence shows the existence of such a fundamental
domain.

Let now $F$ be any fundamental
domain for $\worldsheet$ in $\worldsheetCoveringSpace$.
The following observation will
be essential.

\begin{lemma}
The quotient $\overline{\partial F} := \partial F/\worldsheetInvolution$ is
a 1-dimensional oriented closed submanifold of $\worldsheet$.
\end{lemma}

\proof
We act with $\worldsheetInvolution$
on property (i) of the fundamental domain $F$:
\begin{equation}
\worldsheetInvolution ( \partial
F  ) = \worldsheetInvolution
( F \cap \worldsheetInvolution(F))
= F \cap \worldsheetInvolution(F)
= \partial F
\end{equation}
This shows that $\worldsheetInvolution$ restricts
to an involution on $\partial F$. Since $\worldsheetInvolution$ acts on $\worldsheetCoveringSpace$ without fixed points, the quotient $\partial
F / \worldsheetInvolution$ is a submanifold
of $\worldsheet$, and as $\partial F$ is closed,
so is the quotient. The orientation of $\worldsheetCoveringSpace$
induces an orientation on $F$.  Because $\worldsheetInvolution$
is orientation-reversing, the orientation
of $\worldsheetInvolution(F)$ is opposite
to the one induced on $\worldsheetInvolution(F)$
as a submanifold of $\worldsheetCoveringSpace$.
Hence, $\partial F$ and $\partial(\worldsheetInvolution(F))$
are equal as sets as well as as oriented submanifolds. Thus $\worldsheetInvolution$
preserves the orientation on $\partial
F$. 
\endofproof

\begin{figure}[h]
\begin{center}
\includegraphics{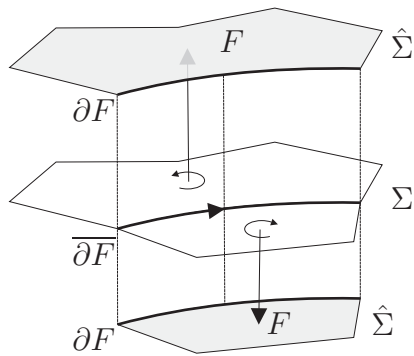}\setlength{\unitlength}{1pt}\begin{picture}(0,0)(276,658)\put(195.83887,773.10992){$\worldsheetCoveringSpace$}\put(195.83887,716.34813){$\worldsheet$}\put(188.63887,668.70992){$\worldsheetCoveringSpace$}\put(149.44394,669.16885){$F$}\put(131.86346,775.25688){$F$}\put(76.81861,662.84135){$\partial F$}\put(76.81861,695.21726){$\overline{\partial F}$}\put(76.81861,749.24135){$\partial F$}\end{picture}
\caption{The orientation on $\overline{\partial
F}$.} 
\end{center}
\end{figure}

\subsection{Unoriented Surface Holonomy}
\label{ss_unsh}

The setup for the definition of
holonomy around closed  unoriented surfaces is
\begin{itemize}
\item a gerbe $\gerbe$ over a smooth manifold $\targetspace$
with Jandl structure $\jandl=(k,\mathcal{A},\varphi)$

\item a closed surface $\worldsheet$

\item a  map $\hat \stringmap \in C^{\infty}(\worldsheetCoveringSpace,M)^{\worldsheetInvolution,k}$
\end{itemize}
 
The idea of the definition is the following: Pull back the gerbe $\gerbe$  to $\worldsheetCoveringSpace$ along $\hat \stringmap$, choose
a trivialization and determine the 2-form
$\hat \rho \in \Omega^2(\worldsheetCoveringSpace)$
as in Definition \ref{hol_bg_or}. Choose a fundamental domain $F$ for $\worldsheet$ in $\worldsheetCoveringSpace$.
The integral
\begin{equation}
\label{bg_hol_try}
\exp \im \int_F \hat \rho
\end{equation}
is independent neither of the choice of
the trivialization -- which enters in $\hat
\rho$ -- nor of the choice of the
fundamental domain $F$. The Jandl structure,
however, allows to correct (\ref{bg_hol_try})
by a boundary term in such a way that the
holonomy becomes well-defined.

\vspace{3mm}

We will now give a detailed definition of
this boundary term, and then show that it
gives rise to a well-defined holonomy. 

Recall that a gerbe $\gerbe$ consist of the following data:
a surjective submersion $\pi:Y \to \targetspace$,
a  line bundle $L \to Y^{[2]}$,  an isomorphism $\mu$, and a
2-form $C \in \Omega^2(Y)$. Recall that the
pullback gerbe $\hat \stringmap^{*}\gerbe$
consists of a pullback
\begin{equation}
\begin{aligned}
\xymatrix{Y_{\stringmap} \ar[r]^{\tilde
\stringmap} \ar[d]_{\pi_{
\stringmap}} & Y \ar[d]^\pi \\ \hat\worldsheet
\ar[r]_{\hat \stringmap} & \targetspace}
\end{aligned}
\text{\;\;,}
\end{equation}
the pullback line bundle $\tilde \stringmap^{*}L$,
isomorphism $\tilde \stringmap^{*}\mu$ and
2-form $\tilde \stringmap^{*}C$. Accordingly,
a trivialization $\mathcal{T}$ of $\hat \stringmap^{*}\gerbe$ is a  line bundle $T \to Y_{\stringmap}$
  together with an isomorphism
\begin{equation}
\tau \maps \tilde \stringmap^{*}L \otimes {\pi_{\stringmap}}_{2}^*T \lto {\pi_{\stringmap}}_{1}^*T
\end{equation}
of  line bundles  
over $Y_{\stringmap}^{[2]}$. It determines
a 2-form $\hat \rho \in \Omega^2(\worldsheetCoveringSpace)$
with \begin{equation}\pi_{\stringmap}^{*}\hat
\rho = \tilde \stringmap ^{*} C - \mathrm{curv}(T)\text{.}\label{rhocurv}\end{equation}

Due to the commutativity of diagram (\ref{bg_hol_co}),
$\hat \stringmap^{*}\mathcal{J}=(\worldsheetInvolution,
\tilde \stringmap^{*}\mathcal{A}, \tilde \stringmap^{*}\varphi)$ is a Jandl structure
on $\hat \stringmap^{*}\gerbe$. Recall that
part of the data are  a  line bundle $\tilde
\stringmap^{*}A \to Z_{\stringmap}$ over
the space $Z_\stringmap := (Y_\stringmap)_\sigma \times_{\worldsheetCoveringSpace}
Y_\stringmap$, and an isomorphism
\begin{equation}
\tilde\stringmap^{*}\alpha \maps p'^*\tilde\stringmap^{*}L \otimes p^*\tilde\stringmap^{*}L^* \otimes {\pi_{\stringmap}}_2^*\tilde\stringmap^{*}A \lto {\pi_{\stringmap}}_1^*\tilde\stringmap^{*}A
\end{equation}
of  line bundles  
over $Z_{\stringmap}^{[2]}$, where
$p$ and $p'$ are the projections in
\begin{equation}
\begin{aligned}
\xymatrix{Z_{\stringmap} \ar[r]^{p}
\ar[d]_{p'} & Y_{\stringmap}
\ar[d]^{\pi_{\stringmap}} \\
{Y_{\stringmap}}_\worldsheetInvolution \ar[r]_{\worldsheetInvolution
\circ \pi_\stringmap} & \worldsheetCoveringSpace}
\end{aligned}\;
\text{ .}
\label{ZYS}\end{equation}
Further, the action of $\orbifoldGroup$
by $\worldsheetInvolution$
lifts to $Z_\stringmap$ via the
permutation map $\tilde \worldsheetInvolution$,
and $\hat \stringmap^{*}\mathcal{J}$
contains an $\orbifoldGroup$-equivariant
structure $\tilde \stringmap^{*}\varphi$
on $\tilde \stringmap^{*}A$. 

Combining the trivialization with the Jandl
structure, we define a  line bundle
\begin{equation}
R := \tilde\stringmap^{*}A \otimes p'^{*}T^{*} \otimes p^{*}T^{*}
\label{bg_hol_te}
\end{equation}
  over $Z_{\stringmap}$.
In addition, we define an isomorphism
\begin{equation}
r:=\tilde\stringmap^{*}\alpha^{-1} \otimes p'^{*}\tau^{*}
\otimes p^{*}\tau^{*} \maps {\pi_\stringmap}_1^{*}R \lto {\pi_\stringmap}_2^{*}R
\end{equation}
of line bundles over $Z^{[2]}_\stringmap$. The compatibility of $\tau$ and $\alpha$
with the isomorphism $\mu$ of $\gerbe$ guarantees
the cocycle condition
\begin{equation}
{\pi_\stringmap}_{23}^*r \circ {\pi_\stringmap}_{12}^*r ={\pi_\stringmap}_{13}^*r
\end{equation}
over $Z^{[3]}_\stringmap$, 
hence $R$ determines a unique  
descent line
bundle $\hat R \to \worldsheetCoveringSpace$,
together with an isomorphism 
${\pi_Z}_\stringmap^{*}\hat R \to R$. We shall compute
the curvature of these bundles, namely
\begin{eqnarray}
\mathrm{curv}\left( R\right) & \stackrel{(\ref{bg_hol_te})}{=} &\tilde{\stringmap}^{*}\mathrm{curv}\left( A\right)
-p'^{*}\mathrm{curv}\left( T\right) -p^{*}\mathrm{curv}
\left( T\right) \\
&\stackrel{(\ref{bg_si_curv})}{=}&p'^{*} (\tilde{\phi}^{*}C-\mathrm{curv}( T))+p^{*}(\tilde{\stringmap}^{*}C -\mathrm{curv}(
T)) \\
&\stackrel{(\ref{rhocurv})}{=}& p'^{*} \pi_\stringmap^{*} \hat \rho+p^{*}\pi_\stringmap^{*} \hat \rho \\ &\stackrel{(\ref{ZYS})}{=}& \pi_{Z_\stringmap}  ^{*}(\worldsheetInvolution^{*}\hat\rho
+ \hat\rho )\text{.}
\end{eqnarray}
Hence the curvature of $\hat R$ is
\begin{equation}
\mathrm{curv}(\hat R)= \worldsheetInvolution^{*}\hat\rho
+ \hat\rho \text{.}
\end{equation}

The next step is to define $\worldsheetInvolution$-equivariant
structure on $\hat R$. Note that the canonical
permutation of tensor products is an equivariant
structure on $p'^{*}T^{*} \otimes p^{*}T^{*}$, since the permutation map $\tilde \worldsheetInvolution$ exchanges $p$ and $p'$.
Together with the equivariant structure $\tilde
\stringmap^{*}\varphi$ on $\tilde \stringmap^{*}
A$, the tensor product (\ref{bg_hol_te})
is the tensor product of two equivariant line
bundles. By definition of
a Jandl structure $\varphi$ is compatible
with $\alpha$, which means that the descent
isomorphism
$r$ is an isomorphism of equivariant 
line bundles. Hence, also
the descent bundle $\hat R$ over $\worldsheetCoveringSpace$ is endowed with
an equivariant structure.

It is a standard fact \cite{gomi1,brylinski2},
that if $K$ is discrete and acts freely,
a $\orbifoldGroup$-equivariant 
line bundle $\hat R \to \worldsheetCoveringSpace$   defines a unique  line bundle $Q$ 
 on the quotient $\worldsheetCoveringSpace
/ \orbifoldGroup = \worldsheet$.

Now choose a fundamental domain $F$ of $\worldsheet$
in $\worldsheetCoveringSpace$.
 
\begin{definition}
\label{hol_bg_def}
The holonomy of the gerbe $\gerbe$
with Jandl structure $\mathcal{J}$
around the unoriented closed surface
$\worldsheet$ is defined as 
\begin{equation}
\label{hol_bg_df}
\mathrm{hol}_{\gerbe,\jandl}(\hat \stringmap,\worldsheet)
:= \exp \left ( \im \int_F \hat \rho \right ) \cdot \mathrm{hol}_Q
(\overline{\partial F})^{-1}\text{.}
\end{equation}
\end{definition} 
In this definition, the compensating term $\mathrm{hol}_Q
(\overline{\partial F})$ is the
holonomy of the  line
bundle $Q$   around
the one-dimensional closed oriented
submanifold $\overline{\partial F}$.

\begin{theorem}
\label{hol_bg_in}
The holonomy defined in Definition
\ref{hol_bg_def} depends neither
on the choice of the fundamental
domain $F$ nor on the choice of
the trivialization $\mathcal{T}$.

\end{theorem}

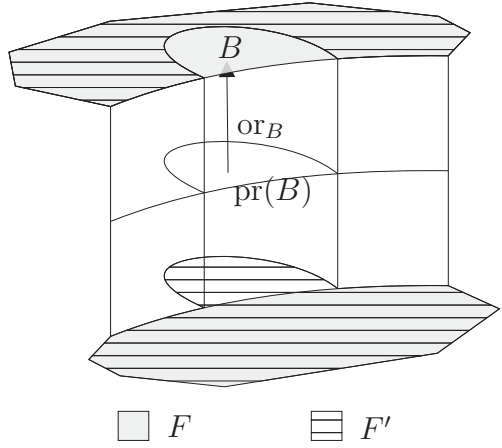
\begin{figure}[h]
\begin{center}
\includegraphics{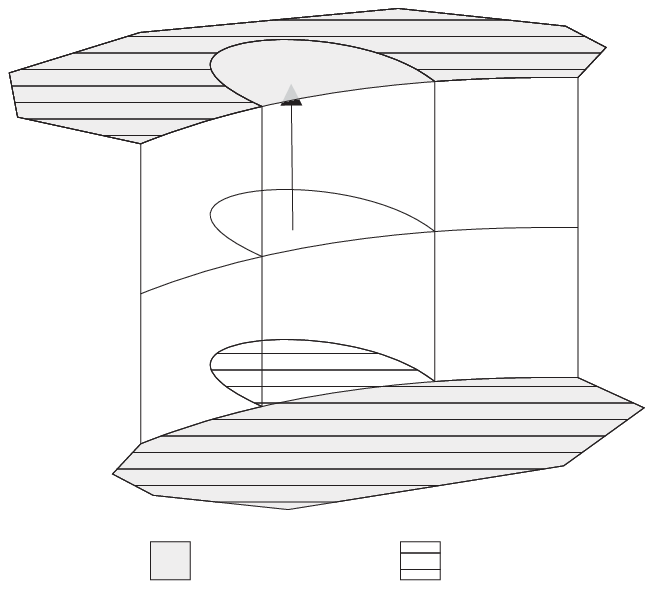}\setlength{\unitlength}{1pt}\begin{picture}(0,0)(223,862)\put(122.56498,953.54447){$\worldsheetCovering(B)$}\put(97.75644,864.10488){$F$}\put(169.75644,863.26668){$F^{\prime}$}\put(116.63263,1007.25817){$B$}\put(124.02425,978.54265){$\mathrm{or}_B$}\end{picture}
\caption{The difference between
two fundamental domains.}
\label{picB}
\end{center}
\end{figure}

\proof
Let $F^{\prime }$ be another fundamental
domain. We define the set
\begin{equation}
B:=\mathrm{Int}(F) \cap \worldsheetInvolution (\mathrm{Int}(F'))\text{,}
\end{equation}
where $\mathrm{Int}$ denotes the interior.
As the intersection of two open sets, $B$ is
open and hence a submanifold of $\worldsheetCoveringSpace$.
It contains those parts of $F$, which are
not contained in $F'$ (cf. Figure
\ref{picB}). Because we excluded
the boundaries of $F$ and $F'$, we have
\begin{equation}
B \cap \worldsheetInvolution (B) = \emptyset \text{,}
\end{equation}
such that there is a unique section  $\mathrm{or}_{B}:\worldsheetCovering (B)
\to \worldsheetCoveringSpace$ with image $B$. 

From Figure \ref{picB}, we have  
\begin{equation}
\int_{F'}\hat{\rho}=\int_{F}\hat{\rho}-\int_{B}\hat{\rho}
+\int_{\worldsheetInvolution (B) }\hat{\rho}=\int_{F}\hat{\rho}-\int_{B}\mathrm{
curv}(\hat{R}) \text{,}
\end{equation}
since $\worldsheetInvolution$ is
orientation-reversing. By Stoke's theorem,
the exponential of the integral of the curvature
of $\hat R$ over $B$ is nothing but the holonomy
of that line bundle around $\partial B$.
Thus,
\begin{equation*}
\exp \left( -\im\int_{B} \mathrm{curv} (\hat{R}) \right) =\mathrm{hol}_{ \hat{R} } ( \partial B) ^{-1}= \mathrm{hol}_{Q} (
\worldsheetCovering (\partial B)) ^{-1}\text{.}
\end{equation*}

This is the term which is  compensated
by the boundary term, which is
\begin{equation}
\mathrm{hol}_{Q} (\overline{\partial F'})^{-1} =\mathrm{hol}_{Q} (\overline{\partial F})^{-1} \cdot \mathrm{hol}_{Q} (\worldsheetCovering(
 \partial B) ) ^{}\text{.}
\end{equation}
In summary
\begin{equation}
\exp \left ( \im \int_{F'} \hat \rho \right ) \cdot \mathrm{hol}_Q
(\overline{\partial F'})^{-1} = \exp \left ( \im \int_F \hat \rho \right ) \cdot \mathrm{hol}_Q
(\overline{\partial F})^{-1}\text{,}
\end{equation}
i.e. the holonomy is independent of the choice
of the fundamental domain. 

Now let $\mathcal{T}'=(\tau' ,T') $ be another
trivialization of $\hat \stringmap ^{*} \gerbe$.
As discussed in section \ref{ss_bg_bg}, there
is a line bundle $N \to \worldsheetCoveringSpace$
together with an isomorphism $\nu : \pi_\phi^{*}N \otimes T' \to T $, such that the 2-forms $\hat \rho$ and $\hat
\rho'$
are related by 
\begin{equation}
\hat{\rho}'=\hat{\rho}+\mathrm{curv}(N) \text{.}
\label{proof1}
\end{equation}
For the line bundle $\hat R$ defined in (\ref{bg_hol_te})
this means
\begin{equation}
R'=R\otimes \pi _{Z}^{*} \worldsheetInvolution ^{*} N\otimes \pi _{Z}^{*} N \text{,}
\end{equation}
and its descent line bundle $\hat R'$ is
\begin{equation}
\hat{R}'=\hat{R} \otimes \worldsheetInvolution ^{*}N\otimes N\text{.}
\label{rrprime}\end{equation}
This is an equation of $\worldsheetInvolution$-equivariant line bundles,
where $\hat R$ and $\hat R'$ obtain equivariant
structures from the Jandl structure as described
before, and  $K:=\worldsheetInvolution
^{*}N \otimes N$ carries the canonical $\worldsheetInvolution $-equivariant structure by permuting the
order in  the
tensor product. Hence, equation (\ref{rrprime})
pushes into the quotient, namely     
\begin{equation}
Q'= Q\otimes \bar{K}\text{.}
\end{equation}
The holonomy of the descent bundle $\bar K$
satisfies
\begin{equation}
\mathrm{hol}_{\bar{K}}( \overline{\partial F}) =\mathrm{hol}
_{N}( \partial F) =\mathrm{hol}_{\worldsheetInvolution ^{*} N} ( \partial F) \text{.}
\label{proof2}\end{equation}
This finally means
\begin{eqnarray}
&&\hspace{-10mm}\exp \left( \im \int_{F}\hat{\rho}'\right) \cdot \mathrm{hol} _ {Q'} (\overline {\partial F}) ^{-1}\nonumber\\
&&\hspace{7mm}\stackrel{(\ref{proof1})}{\hspace{3mm}=} \; \exp \left( 
\im \int_{F} \hat{\rho} + \mathrm{curv} (N) \right) \cdot \mathrm{hol}_{Q \otimes \bar{K}}
(\overline{\partial F} ) ^{-1} \\
&&\hspace{7mm}\stackrel{(\ref{proof2})}{\hspace{3mm}=}\;\exp \left( \im \int_{F}\hat{\rho}\right) \cdot \mathrm{hol}_{N}(\partial F) \cdot \mathrm{hol}_{N} ( \partial F) ^{-1} \cdot \mathrm{hol}_{Q} ( \overline{\partial F}) ^{-1} \\
&&\hspace{7mm}\stackrel{}{\hspace{3mm}=}\;\exp \left( \im \int_{F}\hat{\rho}\right) \cdot \mathrm{hol}_{Q} (\overline{\partial F} ) ^{-1}\text{,}
\end{eqnarray}
thus the holonomy is independent of the choice
of the trivialization.
\endofproof

The following Lemma asserts that
the definition of holonomy is compatible
with the definition of equivalence
of Jandl structures. 

\begin{lemma}
The holonomy of a gerbe $\gerbe$ with Jandl
structure $\jandl$ only depends on the equivalence
class of $\jandl$. 
\end{lemma}

\proof
Let $\jandl=(\orbifoldGroupElement,\mathcal{A},\varphi)$ and $\jandl'=(\orbifoldGroupElement,\mathcal{A}',\varphi')$ be two equivalent
Jandl structures on $\gerbe$. It is shown
in Theorem \ref{bg_js_to} that there is
a unique flat equivariant line bundle $N$
on $\targetspace$,
such that $N.{A} \cong {A}'$
as equivariant line bundles. Because the
action of $\mathrm{Pic}_{0}^{\orbifoldGroup}(\targetspace)$
is free, and $A$ and $A'$ are isomorphic, $N$ is the trivial equivariant line
bundle. Remember the
definition of the bundle $R\to Z$ in
equation (\ref{bg_hol_te}). For the two Jandl
structures we get $R'=R \otimes \pi_Z^{*}N$,
and hence the descent bundles $\hat R' = \hat
R \otimes N$ over $\worldsheetCoveringSpace$. Since $N$ is the trivial equivariant
line bundle, $\hat R'$ and $\hat R$ are isomorphic
as equivariant line bundles, and thus define
isomorphic line bundles $Q'$ and $Q$ over
$\worldsheet$. Isomorphic line bundles have
the same holonomies, so Definition \ref{hol_bg_def}
is independent of the equivalence class of
$\jandl$. \endofproof

An important condition for any notion of
unoriented surface holonomy is its compatibility
with ordinary surface holonomy for oriented
surfaces: 

\begin{theorem}
\label{hol_bg_red}
If $\worldsheet$ is orientable, for any choice
of an orientation, the
holonomy defined in Definition
\ref{hol_bg_def} reduces to the
ordinary holonomy defined in Definition
\ref{hol_bg_or}, 
\begin{equation}
\mathrm{hol}_{\gerbe,\jandl}(\hat\stringmap,\worldsheet)
= \mathrm{hol}_\gerbe(\stringmap,
\worldsheet)\text{,}
\end{equation}
where $\stringmap$ and $\hat\stringmap$
are related by
the bijection of Lemma \ref{hol_bg_bi}. In
particular, if $\gerbe$ admits a Jandl structure,
the holonomy of $\gerbe$ does not depend
on the orientation. 
\end{theorem}

\proof
Let $\mathrm{or}:\worldsheet\to\worldsheetCoveringSpace$
be a choice of an orientation on $\worldsheet$.
Then $F := \mathrm{or}(\worldsheet)$
is a fundamental domain with empty
boundary $\partial F=\emptyset$.
Choose a trivialization $\mathcal{T}$
of $\hat\stringmap^{*}\gerbe$ to obtain the 2-form $\hat\rho\in\Omega^2(\worldsheetCoveringSpace)$.
Then the left hand side is equal
to $\exp\im\int_{\mathrm{or}(\worldsheet)} \hat\rho$, because of Theorem \ref{hol_bg_in}.
Because $\hat\stringmap$ and $\stringmap$
correspond to each other, $\mathrm{or}^{*}\hat\stringmap^{*}\gerbe$
is the same gerbe as $\stringmap^{*}\gerbe$,
and $\mathrm{or}^{*}\mathcal{T}$
is a trivialization with 2-form
$\rho = \mathrm{or}^{*}\hat\rho$.
Thus, the right hand side is equal
to $\exp\im\int_\worldsheet\rho$ and therefore
equals the ordinary holonomy.
\endofproof

\subsection{Holonomy in Local Data}
\label{ss_hol_ld}

Let $\left \lbrace \targetspaceCover_i \right \rbrace_{i \in I}$ be an
 open cover of $\targetspace$.
 To avoid notation, we assume that
 it is invariant under $\orbifoldGroupElement$
 and still good enough to admit
 all the local sections necessary to
 extract  local data
$(\gerbelocijk,\gerbelocij,\gerbeloci)$
of the gerbe $\gerbe$ and $(\jandllocij,\jandlloci,\jandlloccc)$
of the Jandl structure $\jandl$,
as we explained in section \ref{ss_ld}.
We pull back the cover $\left \lbrace \targetspaceCover_i \right \rbrace_{i \in I}$ along  $\hat\stringmap:\worldsheetCoveringSpace\to\targetspace$
 and obtain a cover $ \lbrace \hat\worldsheetCover_i  \rbrace_{i \in I}$ with $\hat\worldsheetCover_i:=\hat\stringmap^{-1}(\targetspaceCover_i)$,
together with  pullback local
data. Next, choose local data $(\gerbetrivij,
\gerbetrivi)$ of the trivialization $\mathcal{T}$ of the pullback gerbe and a 2-form $\hat\rho\in\Omega^2(\worldsheetCoveringSpace)$, so that
\begin{equation}
\label{hol_ld_r}
\left(\hat\stringmap^{*}\gerbelocijk,\hat\stringmap^{*}\gerbelocij,\hat\stringmap^{*}\gerbeloci
\right) =\left( 1,0,\hat\rho \right) +\deligneDifferential\left( \gerbetrivij,
\gerbetrivi
\right)
\end{equation}
holds. Following the definition
of the bundle $R\to Z$ in equation (\ref{bg_hol_te}), the 
bundle $\hat R\to\worldsheetCoveringSpace$
has local data
\begin{equation}
(r,R) := \hat \stringmap^{*}(\jandllocij,
\jandlloci)-\worldsheetInvolution^{*}(\gerbetrivij,\gerbetrivi)-
(\gerbetrivij,\gerbetrivi) \text{;}
\end{equation}
the condition that $\hat R$ descends is equivalent to the
Deligne cocycle condition
\begin{equation}
\deligneDifferential(r,R)=(1,0)\text{,} 
\end{equation}
which follows from equations (\ref{hol_ld_r})
and (\ref{bg_ld_c1}).

Because $\hat\stringmap$ is an element of
$C^{\infty}(\worldsheetCoveringSpace,M)^{\worldsheetInvolution,\orbifoldGroupElement}$,
the pullback cover is invariant
under $\worldsheetInvolution$.
Hence it projects to a cover of
$\worldsheet$ with
open sets
$\worldsheetCover_i:=\worldsheetCovering(\hat\worldsheetCover_i)$.
Choose local sections $\mathrm{or}_i:\worldsheetCover_i\to\worldsheetCoveringSpace$
and a dual triangulation $T$ of $\worldsheet$, subordinate
to  the cover $\lbrace \worldsheetCover_i \rbrace_{i \in
I}$, together with a subordinating map $i:
T \to I$. As we did in section
\ref{ss_fd} we choose the fundamental
domain 
\begin{equation}
F := \bigcup_{f \in T} \mathrm{or}_{i(f)}(f)\text{,}
\end{equation}
where the $f$'s are the faces of the triangulation.

We now introduce  three abbreviations. Let
$\omega_{i}^{2} \in \Omega^2(\hat\worldsheetCover_i)$,
$\omega_{ij}^{1} \in \Omega^1(\hat\worldsheetCover_i
\cap \hat\worldsheetCover_j)$ and
$\omega_{ijk}: \hat\worldsheetCover_i \cap
\hat\worldsheetCover_j  \cap \hat\worldsheetCover_k
\to U(1)$ be some  local data. 
First we denote the integral over a face $f$ by
\begin{eqnarray}
\nonumber I_{f }(\omega,
\omega^1 , \omega^2) &:=& \exp \left( \im \int_{ \mathrm{or}
_{i(f) }( f) } \omega^2_{i( f) }+ 
\im \sum_{e \in \partial f} \int_{\mathrm{or}_{i
(f) } ( e ) } \omega^1 _ {i ( f ) i ( e ) } \right)
\\ &&\cdot \prod _ {v \in \partial e} \omega
_ { i ( f ) i ( e ) i ( v ) } ^ {\varepsilon ( f ,e,v ) } ( \mathrm{or} _ {i ( f ) } (  v ) ) \text{,} 
\end{eqnarray}
where $\varepsilon ( f ,e,v ) \in \left\{ 1,-1\right\} $ indicates, whether $v$ is the end
or the starting point of the edge $e$ with
respect to the orientation  $%
\mathrm{or}_{i ( f ) }$. 

Second, we  denote
the integral of some local data $\omega_{i}^1\in\Omega^1(\hat U_i)$
and $\omega_{ij}:\hat U_i \cap \hat U_j \to U(1)$ along an edge $e$ of a face
$f$ by
\begin{equation}
I_{e,f }( \omega, \omega^1 ) :=\exp \left( \im \int_{\mathrm{or}_{i ( f ) } ( e ) } \omega^1_{i ( e ) } \right) \cdot
\prod _ { v \in \partial e} \omega_{i
( e ) i ( v ) }^{\varepsilon ( f ,e,v ) }
( \mathrm{or}_{i ( f ) } ( v ) ) \text{.}
\end{equation} Recall that the set of edges in $T$ splits
into the set $E$ of orientation-preserving
edges and the set $\bar E$ of orientation-reversing
edges. For an orientation-preserving
edge $e \in f_1 \cap f_2$ we have 
\begin{equation}
\label{hol_ld_oc}
I_{e,f_1 }( \omega, \omega^1 ) =I_{e,f_2 }( \omega, \omega^1 ) ^{-1}%
\text{,}
\end{equation}  
while for an orientation-reversing edge
\begin{equation}
I_{e,f_1 }( \omega, \omega^1 )
= I_{e,f_2 }( \worldsheetInvolution^{*} \omega, \worldsheetInvolution^{*} \omega^1 )\\
\end{equation}
holds. In the latter case, since $e$ is orientation-reversing,
we have either $\mathrm{or}_{i(e)}(e)=\mathrm{or}_{i(f_1)}(e)$
or $\mathrm{or}_{i(e)}(e)=\mathrm{or}_{i(f_2)}(e)$,
so that we can write just $I_{e}( \omega, \omega^1 )$, where the for $f$ the choice
of the face with the coinciding
orientation is understood.

Third, if $v$ is a vertex of an edge $e$,
we define for some smooth function $\omega_i
: \hat U_i \to U(1)$
\begin{equation}
I_{v,e, f } ( \omega ) := \omega_{i ( v ) }^{\varepsilon ( f ,e,v ) } ( \mathrm{or}_{i( f ) } ( v )  )\text{.}
\end{equation}
Now if $v$
is the common vertex of two orientation-reversing
edges $e_1,e_2\in\bar E$, we call $v$ orientation-preserving,
if $\mathrm{or}_{i(e_1)}(v) = \mathrm{or}_{i(e_2)}(v)$
and orientation-reversing otherwise. Let
us denote the set of orientation-reversing
vertices by $\bar V$. If $v$ is such a vertex,
we just write $I_v(\omega)$ instead of $I_{v,e, f } ( \omega )$, where for $e$ the choice of the edge as well as for $f$  the
face with the coinciding orientation is understood.

Now the first factor in the holonomy formula (\ref{hol_bg_df}) is
\begin{equation}
\exp \left( \im \int_{F} \hat\rho \right)
= \exp \left( \im \sum_{f \in T} \int_{\mathrm{or}_{i(f)}(f)} \hat\stringmap^{*}B_{i(f)}  + \exteriorDifferential
\gerbetrivi_{i(f)} \right)\text{.}
\end{equation}
Following \cite{carey2}, by using Stoke's theorem, equation
(\ref{hol_ld_r}) and our abbreviations, we
end up with
\begin{equation}
\label{nachstokes}
\exp \left( \im \int_{F} \hat\rho \right)
=\prod _{f \in T} I_{f} ( \hat\stringmap
^* \gerbelocijk ,\hat\stringmap^*\gerbelocij,
\hat\stringmap^*\gerbeloci ) \cdot \prod_{f
\in T}\prod _{e \in \partial f } I_ {e,f } ( \gerbetrivij , \gerbetrivi) ^{-1} \text{.}
\end{equation}
Here the second factor collects the boundary
contributions that appear in the application
of Stoke's theorem. 

Let us assume for the moment that $\worldsheet$
the oriented, and all sections $\mathrm{or}_i$
coincide with the global orientation restricted
to $\worldsheetCover_i$. In this situation,
we have only orientation preserving edges,
and each of them appears twice in the second factor. Since the contributions are inverse
by (\ref{hol_ld_oc}), the second factor vanishes.
We obtain the local holonomy formula expressed
only by the local data of the gerbe, as it
appeared originally in \cite{alvarez1}.

If $\worldsheet$ is not oriented, the second
factor still consists of two
contributions for each orientation-reversing
 edge $e\in \bar E$, which are 
\begin{equation}
I_{e,f_1} ( \gerbetrivij ,\gerbetrivi) \cdot I_{e,f_2} ( \gerbetrivij , \gerbetrivi) =
I_{e} ( \gerbetrivij \cdot \worldsheetInvolution ^*\gerbetrivij , \gerbetrivi + \worldsheetInvolution ^* \gerbetrivi )\text{.}
\end{equation} 
Hence, in the general case, the second factor
of (\ref{nachstokes})
is
\begin{equation}
\label{hol_ld_sf}
 \prod_{f
\in T}\prod _{e \in \partial f } I_ {e,f } ( \gerbetrivij , \gerbetrivi) ^{-1} = \prod_{e
\in \bar E} I_{e} ( \gerbetrivij \cdot \worldsheetInvolution ^*\gerbetrivij , \gerbetrivi + \worldsheetInvolution ^* \gerbetrivi )^{-1}\text{.}
\end{equation}
   
\begin{figure}[h]
\begin{center}
\includegraphics{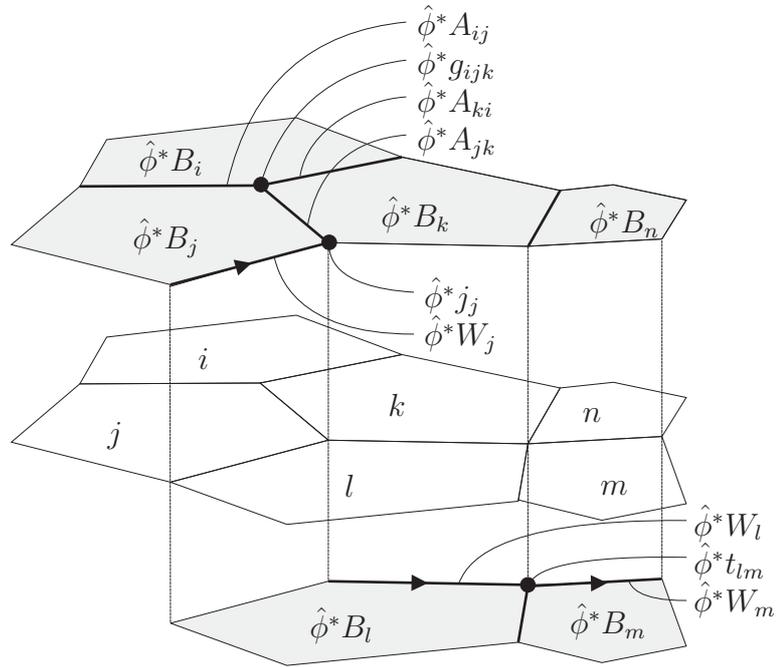}\setlength{\unitlength}{1pt}\begin{picture}(0,0)(645,332)\put(355.80586,446.25430){$i$}\put(321.90576,418.35487){$j$}\put(426.84718,428.61317){$k$}\put(499.67773,426.58809){$n$}\put(410.63641,398.09877){$l$}\put(541.20898,383.70529){$\hat \stringmap^{*}\jandlloci_{l}$}\put(397.53808,345.10649){$\hat \stringmap^{*}\gerbeloci_{l}$}\put(437.64917,529.12431){$\hat \stringmap^{*}\gerbelocij_{jk}$}\put(333.60605,521.18787){$\hat \stringmap^{*}\gerbeloci_{i}$}\put(331.30658,491.69055){$\hat \stringmap^{*}\gerbeloci_{j}$}\put(506.35276,398.93443){$m$}\put(502.14869,497.80602){$\hat \stringmap^{*}\gerbeloci_n$}\put(494.92998,346.64088){$\hat \stringmap^{*}\gerbeloci_m$}\put(541.20898,369.30529){$\hat \stringmap^{*}\jandllocij_{lm}$}\put(541.20898,354.90529){$\hat \stringmap^{*}\jandlloci_{m}$}\put(440.40898,455.27386){$\hat \stringmap^{*}\jandlloci_{j}$}\put(440.40898,469.67386){$\hat \stringmap^{*}\jandlloccc_{j}$}\put(437.64917,543.52431){$\hat \stringmap^{*}\gerbelocij_{ki}$}\put(437.64917,557.92431){$\hat \stringmap^{*}\gerbelocijk_{ijk}$}\put(437.64917,572.32431){$\hat \stringmap^{*}\gerbelocij_{ij}$}\put(424.38671,500.04170){$\hat \stringmap^{*}\gerbeloci_k$}\end{picture}
\caption{Assignment of local data. The middle
layer shows $\worldsheet$ and the subordinated
indices; the top and lower layer show parts
of the two  sheets of $\worldsheetCoveringSpace$.}
\label{picDaten}
\end{center}
\end{figure}

For the second factor of the holonomy formula
(\ref{hol_bg_df}) we have to compute the
holonomy of the descent line bundle $Q$ around $\overline{\partial
F}$. Note that 
\begin{equation}
\hat{\bar E} := \bigcup_{e \in \bar E}\mathrm{or}_{i(e)}(e)
\end{equation}
is a fundamental domain of $\overline{\partial
F}$ in $\partial F$ with boundary consisting
of the preimages of the orientation-reversing
vertices $v \in \bar V$. Now the holonomy
of $Q$ around $\overline{\partial F}$ is
equal to the the holonomy of $\hat R$ around $\hat{\bar E}$, where at the boundary points the equivariant
structure of $\hat R$ is used, this is
\begin{equation}
\mathrm{hol}_{Q} ( \overline{\partial F}
) = \prod _{e\in \bar{E}}I_{e} ( r,R ) \cdot \prod _{v \in \bar{V}} I_{v} ( 
\hat{\stringmap}^* \jandlloccc)\text{.}
\end{equation}
Since $e$ is orientation-reversing, 
\begin{eqnarray}
I_{e} ( r,R ) &=& I_{e} ( \hat{\stringmap}^*
\jandllocij \cdot \worldsheetInvolution
^* \gerbetrivij ^{-1}\cdot \gerbetrivij ^{-1},\hat{\stringmap}^*
\jandlloci -\worldsheetInvolution ^* \gerbetrivi
- \gerbetrivi) \\
\label{hol_ld_ff}
&=& I_{e} ( \hat{\stringmap}^* \jandllocij ,\hat{\stringmap}^*
\jandlloci ) \cdot
I_{e} ( \gerbetrivij \cdot \worldsheetInvolution ^* \gerbetrivij , \gerbetrivi + \worldsheetInvolution ^* \gerbetrivi ) ^{-1}\text{.}
\end{eqnarray}
The second factor of (\ref{hol_ld_ff}) cancels
(\ref{hol_ld_sf}) so that all the local data coming from the trivialization drops
out. It remains
\begin{equation}
\mathrm{hol}_{\gerbe,\jandl} ( \worldsheet,
\hat \stringmap)
= \prod_{f \in T} I_{f} ( \hat{\stringmap}^*
\gerbelocijk ,\hat{\stringmap}^*\gerbelocij,\hat{\stringmap}^*\gerbeloci
) \cdot \prod_{e\in \bar{E} }I_{e} ( \hat{\stringmap}^*
\jandllocij ,\hat{\stringmap}^* \jandlloci
) ^{-1} \cdot \prod_{v\in \bar{V}} I_{v}
( \hat{\stringmap}^* \jandlloccc ) 
\text{,}
\label{hol_ld}\end{equation}
depending only on the local data
of the gerbe and of the Jandl structure.
We visualize this formula in Figure
\ref{picDaten}. 

\subsection{Examples}

In the next two subsections we
will apply the general formula (\ref{hol_ld})
to some examples of surfaces $\worldsheet$,
and we will simplify the situation
considerably by starting with the pullback
gerbe $\hat \stringmap ^{*} \gerbe$
which allows us to choose a triangulation adapted to $\worldsheet$. 

\subsubsection{Klein Bottle}

\begin{figure}[h]
\begin{center}
\includegraphics{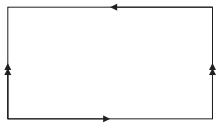}\setlength{\unitlength}{1pt}\begin{picture}(0,0)(135,236)\end{picture}
\caption{Klein Bottle.}
\label{picKleinBottle}
\end{center}
\end{figure}

\begin{figure}[h]
\begin{center}
\includegraphics{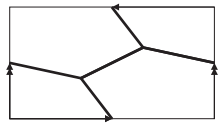}\setlength{\unitlength}{1pt}\begin{picture}(0,0)(225,236)\end{picture}
\caption{Klein Bottle with a dual
\label{picKleinBottleTriang}
triangulation.}
\end{center}
\end{figure}

\begin{figure}[h]
\begin{center}
\includegraphics{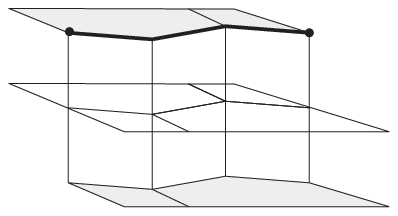}\setlength{\unitlength}{1pt}\begin{picture}(0,0)(476,226)\put(402.99335,283.29307){$F$}\put(428.32403,226.02444){$F$}\put(370.41279,271.62709){$v$}\put(453.57732,270.36312){$v$}\end{picture}
\caption{A fundamental domain for
the Klein Bottle in its double
covering.}
\label{picKleinBottleDouble}
\end{center} 
\end{figure}

Think of the Klein bottle as a
rectangle with the identifications of the
boundary indicated by arrows as
in Figure \ref{picKleinBottle}.
The identification by the vertical arrows is orientation-preserving,
while the one by the 
horizontal arrows is orientation-reversing.
A dual triangulation is shown in
Figure \ref{picKleinBottleTriang}.
Note that this is    a  triangulation
with only one face. We choose a
local section from that face into
the double cover, and define the
fundamental domain $F$ as its image,
as indicated in Figure
\ref{picKleinBottleDouble}.
Here we dropped the arrows, but the
identifications are still to be understood,
so that  both points labelled by $v$ are identified.
This means, that we can choose
the local orientations of the edges
such that the orientation-reversing
edges form a closed line, as
indicated by the thick line. So
there is no orientation-reversing
vertex, and the local datum $\jandlloccc$
of the Jandl structure is not relevant for the holonomy
around the Klein bottle.

\subsubsection{The real projective
Plane}
\label{ss_pp} 

\begin{figure}[h]
\begin{center}
\includegraphics{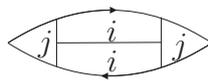}\setlength{\unitlength}{1pt}\begin{picture}(0,0)(225,174)\put(185.15310,188.81858){$i$}\put(185.17833,177.31729){$i$}\put(209.82431,184.04031){$j$}\put(159.42431,185.04031){$j$}\end{picture}
\caption{A dual triangulation of
the real projective plane with
two faces.}
\label{picRP2}
\end{center}
\end{figure}

\begin{figure}[h]
\begin{center}
\includegraphics{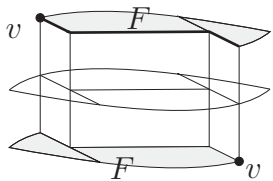}\setlength{\unitlength}{1pt}\begin{picture}(0,0)(479,151)\put(409.96800,151.22750){$F$}\put(416.11011,207.67890){$F$}\put(460.77732,151.56312){$v$}\put(370.98992,204.03950){$v$}\end{picture}
\caption{A fundamental domain of
the real projective plane in its
double covering.}
\label{picRP2double}
\end{center}
\end{figure}

We proceed in the same way as for
the Klein bottle, so think of the
real projective plane $\R P^2$
as
\begin{figure}[h]
\begin{center}
\includegraphics{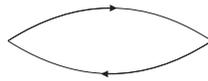}\setlength{\unitlength}{1pt}\begin{picture}(0,0)(135,174)\end{picture}
\caption{The real projective plane.}
\label{picRP2a}
\end{center}
\end{figure}
a two-gon with the identification on the
boundary  indicated by arrows in Figure
\ref{picRP2a}.
The identification is orientation-reversing.
An example of a  dual triangulation is for example
shown in Figure \ref{picRP2}.
Now we choose
local sections from these two faces
into the double cover, for example
as shown in Figure \ref{picRP2double}.
Note that here the thick line is
not a closed line in $\worldsheetCoveringSpace$,
and
$v$ is an orientation-reversing
vertex. According to the local holonomy
formula (\ref{hol_ld}) here the
local datum $\jandlloccc$ of the
Jandl structure enters in the holonomy. 

\section{Gerbes and Jandl Structures in WZW Models}
\label{ss_wzw}

\subsection{Oriented and orientable WZW Models}

In the following we are concerned with Lie
groups $\targetspace$, and  we will use the following notation.
 The left multiplication with a group element
$h$ is denoted by $l_h:\targetspace \to \targetspace$,
and the map which assigns to $h$
 the inverse
group  element $h^{-1}$ is denoted by  $\mathrm{Inv}:\targetspace
\to \targetspace$. 
The left invariant Maurer-Cartan form is
denoted by $\theta$, and the right invariant
form by $\bar \theta$. We call a gerbe $\gerbe$ over $\targetspace$
left invariant, if it is stably isomorphic to
the gerbe $l_h^{*}\gerbe$ for each $h \in
\targetspace$, and similar for right
and bi-invariance.

A WZW model is a theory of maps $\stringmap:
\worldsheet \to \targetspace$ from a
worldsheet $\worldsheet$ into a target space
$\targetspace$, which is a Lie group together
with additional structure,
called the background fields.   It assigns to each 
map $\stringmap$ an  amplitude, i.e. a number in $U(1)$,
as the weight of this map in a path integral.
To be more precise:

\begin{definition}
An oriented WZW model consists of  a compact connected Lie
group $\targetspace$, which is equipped with
an $\mathrm{Ad}$-invariant metric $g=\langle-,-\rangle$ on its Lie algebra
and  a bi-invariant gerbe $\gerbe$. It
assigns an amplitude 
\begin{equation}
A^\text{ortd}_{g,\gerbe}(\stringmap, \worldsheet):=\exp \left( \im S_{\text{kin}}(\stringmap) \right) \cdot \mathrm{hol}_\gerbe
(\worldsheet, \stringmap)
\label{chiralwzw}
\end{equation}
to a map $\stringmap: \worldsheet \to \targetspace$
from a closed oriented conformal worldsheet
$\worldsheet$ to $\targetspace$, where the kinetic term is
\begin{equation}
S_{\text{kin}}(\stringmap) := \frac{1}{2}  \int_\worldsheet
\left\langle \stringmap^{*}\theta
\wedge\star\stringmap^{*}\theta \right\rangle\text{.}
\label{wzwkin}\end{equation}        
\end{definition}
Note that the conformal structure and the
orientation on $\worldsheet$ determine the
Hodge star. 

\vspace{3mm}

In \cite{witten1} Witten discussed
this theory for $\targetspace = SU(2)$, which
is an example for a compact, simple, connected
and simply-connected Lie group.
In this particular situation, the holonomy
can be written as the exponential of the Wess-Zumino term,
\begin{equation}
\mathrm{hol}_\gerbe
(\worldsheet, \stringmap) = \exp \left (
\im \int_B \tilde \stringmap^{*}H\right )\text{,}
\end{equation}
so that we can express the amplitudes
as 
\begin{equation}
A^\text{ortd}_{g,\gerbe}(\stringmap, \worldsheet)=\exp(\im S_{\text{WZW}}(\stringmap))
\end{equation}
with the action functional 
\begin{equation}
S_{\text{WZW}}(\stringmap) := S_{\text{kin}}(\stringmap) + \int_B \tilde \stringmap
^{*} H \text{.}
\end{equation}
Here $B$ is a 3-dimensional manifold with
boundary $\worldsheet$, $\tilde \stringmap$
is an extension of $\stringmap$ on $B$, and
$H$ is the curvature of the gerbe $\gerbe$.   

Witten observed two symmetries of
the WZW model on the type of Lie groups he
considered. The first is translation symmetry:
the action functional
$S_{\text{WZW}}(\stringmap)$ is
invariant under the translation
$\stringmap \mapsto l_h \circ \stringmap$.
 The associated conserved Noether
current is given
by
\begin{equation}
J(\stringmap) := -(1+\star)\stringmap^{*}\theta\text{,}
\end{equation}
which is a 1-form on $\worldsheet$
with values in the Lie algebra
of $\targetspace$. To obtain this conserved, non-abelian current, Witten derived a specific relative
normalization
of the kinetic and the Wess-Zumino term, which was also adapted here. 

The second symmetry Witten observed is the invariance of the action functional $S_{\text{WZW}}(\stringmap)$ under
what he called parity transformation:   reverse the orientation on $\worldsheet$
and replace $\stringmap$ by $\bar \stringmap
:= \mathrm{Inv}
\circ \stringmap$. Accordingly,
the conserved current $J(\stringmap)$ for $\Sigma$ and
the one for $\bar \worldsheet$,
the manifold $\Sigma$ with reversed orientation, namely
\begin{equation}
\bar J (\bar \stringmap)=(1-\star)\stringmap^{*}\bar\theta\text{,}
\end{equation}
are often called equivalent. Note that here
the right invariant Maurer-Cartan form appears.
In that sense, the parity transformation exchanges
left and right movers.

\vspace{3mm}

We now want to generalize this
equivalence to any compact connected Lie group
$\targetspace$. It is a simple
consequence of the properties of
the holonomy of $\gerbe$, that
the parity symmetry
\begin{equation}
A^\text{ortd}_{g,\gerbe}(\stringmap, \worldsheet) = A^\text{ortd}_{g,\gerbe}(\mathrm{Inv}
\circ \stringmap, \bar\worldsheet)
\label{wzwparity}\end{equation}
holds, if the gerbes $\mathrm{Inv}^{*}\gerbe$ and $\gerbe^{*}$
are stably
isomorphic. Note that this is a condition on the  gerbe $\gerbe$.
It should not come as a surprise that in Witten's discussion
there is no such condition: 

\begin{lemma}
\label{paritysu2}
If $\gerbe$ is a bi-invariant gerbe over a
compact, simple, connected and simply connected
Lie group, then $\mathrm{Inv}^{*}\gerbe$ and $\gerbe^{*}$ are stably isomorphic. 
\end{lemma}

\proof Because stably isomorphic gerbes have
the same curvatures, the curvature $H$ of
the bi-invariant gerbe $\gerbe$ is a  bi-invariant 3-form. It is a theorem
by Cartan, that on compact, simple, connected, simply connected
Lie groups $\targetspace$ the space of bi-invariant
3-forms is  the span of the canonical  3-form
$\nu$, which satisfies $\mathrm{Inv}^{*}\nu
= -\nu$. Hence $\mathrm{Inv}^{*}\gerbe$ and $\gerbe^{*}$ have the same curvature. Because the set of stable isomorphism
classes of gerbes of  same curvature
form a torsor over $\mathrm{H}^2(\targetspace,U(1))$
\cite{gawedzki1}, which here is the
trivial group,
the gerbes
$\mathrm{Inv}^{*}\gerbe$ and $\gerbe^{*}$
are stably
isomorphic.
\endofproof

We now give an even more general definition
of parity transformations of  a target space
$\targetspace$ with metric
$g$ and gerbe $\gerbe$.  

\begin{definition}
\label{def_parity}
A parity transformation map is an isometry $\orbifoldGroupElement:\targetspace
\to \targetspace$ of the metric $g$ of order
two, such that
$\orbifoldGroupElement^{*}\gerbe$ and
$\gerbe^{*}$  are stably isomorphic. We denote
the set of parity transformation maps by
$P(M,g,\gerbe)$.
\end{definition}

Consider an oriented WZW
model with target space $\targetspace$, $\mathrm{Ad}$-invariant metric $g$ and bi-invariant gerbe
$\gerbe$. If $\orbifoldGroupElement \in P(M,g,\gerbe)$
is a parity transformation map, we obtain
the parity symmetry
\begin{equation}
A^\text{ortd}_{g,\gerbe}(\stringmap, \worldsheet) = A^\text{ortd}_{g,\gerbe}(\orbifoldGroupElement
\circ \stringmap, \bar\worldsheet)\text{.}
\label{wzwgenpar}\end{equation}

We already discussed that $\orbifoldGroupElement
= \mathrm{Inv}$ is  a parity transformation
map
in the sense of Definition \ref{def_parity},
if the gerbes $\mathrm{Inv}^{*}\gerbe$ and
$\gerbe^{*}$ are stably isomorphic. 
However,  for oriented WZW models on  compact connected Lie groups
there are more such parity transformation
maps. Because the  gerbe
$\gerbe$ is supposed to be bi-invariant,
we try an ansatz $\orbifoldGroupElement := l_h \circ
\mathrm{Inv}$ for some group element
$h \in \targetspace$. The condition
$\orbifoldGroupElement^2 = \mathrm{id}_\targetspace$
restricts $h$ to be an element
of the center $Z(\targetspace)$.  So, the
set $P(\targetspace,g,\gerbe)$ of parity transformation maps for a compact
connected Lie group $\targetspace$ and a
bi-invariant gerbe $\gerbe$, such that
$\gerbe^{*}$ is stably isomorphic
to $\mathrm{Inv}^{*}\gerbe$, contains
at least
\begin{equation}
\left \lbrace l_z \circ \mathrm{Inv} \; |
\; z \in Z(\targetspace) \right \rbrace \subset
P(\targetspace,g,\gerbe)\text{.}
\end{equation}
In particular, $P(\targetspace,g,\gerbe)$
is not empty in the situation we are interested
in.  

\vspace{3mm}

As a preparation for the unoriented case,
we now relate parity symmetry
to the orientation cover $\worldsheetCoveringSpace$:
Start with an oriented WZW model
on $\worldsheet$ together with a parity transformation
map $\orbifoldGroupElement$. Let $\stringmap: \worldsheet \to \targetspace$
be a map.  By Lemma \ref{hol_bg_bi}, there is a unique  map $\hat \stringmap \in
C^{\infty}(\worldsheetCoveringSpace, \targetspace)^{\orbifoldGroupElement,\worldsheetInvolution}$.
 Once we have the orientation cover $\worldsheetCoveringSpace$
and the map $\hat \stringmap$,
we may forget their origin, in particular the
orientation on $\worldsheet$. Then we may
give the following

\begin{definition}
An orientable WZW model consists of a compact connected Lie
group $\targetspace$, which is equipped with
an $\mathrm{Ad}$-invariant metric $g$ on its Lie
algebra, a bi-invariant
gerbe $\gerbe$ and a parity transformation map $\orbifoldGroupElement
\in P(\targetspace,g,\gerbe)$.  To a closed orientable conformal surface
$\worldsheet$ and a map     $\hat \stringmap \in
C^{\infty}(\worldsheetCoveringSpace, \targetspace)^{\orbifoldGroupElement,\worldsheetInvolution}$,
the following amplitude $A^\text{orble}_{g,\gerbe}(\hat \stringmap, \worldsheet)$ is assigned. Choose
any orientation on $\worldsheet$, and obtain
a map $\stringmap: \worldsheet \to \targetspace$
by Lemma \ref{hol_bg_bi}. Define 
\begin{equation}
A^\text{orble}_{g,\gerbe}(\hat \stringmap, \worldsheet) := A^\text{ortd}_{g,\gerbe}(\stringmap, \worldsheet)
\text{.}
\label{wzwamplor}
\end{equation} 
\end{definition}

The amplitude is well-defined: if we had chosen the other orientation,  we would get the same amplitudes, due to
the fact that $\orbifoldGroupElement$ is
a parity transformation map and satisfies
equation (\ref{wzwgenpar}).
  
       
\subsection{Unoriented  WZW Models}
\label{ss_unor}

In the last section we gave the definition
of an orientable WZW model. The derivation
of the amplitude of a map $\hat \stringmap \in
C^{\infty}(\worldsheetCoveringSpace, \targetspace)^{\orbifoldGroupElement,\worldsheetInvolution}$  makes use of the existence
of an orientation on $\worldsheet$ both in
the kinetic term and in the holonomy term.
In this section, we want to overcome this
obstruction. 

\vspace{3mm}  

Let us first discuss the kinetic term. We
want to define the kinetic term $S_{\text{kin}}(\hat \stringmap) $ for a map $\hat \stringmap \in
C^{\infty}(\worldsheetCoveringSpace, \targetspace)^{\orbifoldGroupElement,\worldsheetInvolution}$
in such a way that if $\worldsheet$ is orientable,
it reduces to the kinetic term 
$S_{\text{kin}}( \stringmap)$ of the corresponding
map $\stringmap$. 
Note
that 
\begin{equation}
\mathcal{L}(\hat \stringmap) := \frac{1}{2} \left\langle \hat\stringmap^{*}\theta
\wedge\star\hat\stringmap^{*}\theta \right \rangle
\end{equation}
is a 2-form on $\worldsheetCoveringSpace$,
which satisfies 
\begin{equation}
\worldsheetInvolution^{*}{\mathcal{L}}(\hat \stringmap) = -{\mathcal{L}}(\hat \stringmap)\text{.}
\label{wzwdensity}\end{equation}
This property tells
us that $\mathcal{L}(\hat \stringmap)$ defines
a 2-density $\mathcal{L}_\text{den}(\hat \stringmap)$ \cite{bott1,berger1}   on $\worldsheet$. The integral of a 2-density
  over a surface is defined without respect to the orientability of this surface, so
we define
\begin{equation}
S_{\text{kin}}(\hat \stringmap) := \int_\worldsheet
\mathcal{L}_\text{den}(\hat \stringmap)
\text{.}
\label{intdens}
\end{equation}

To make the integral (\ref{intdens}) more
explicit, choose a triangulation $T$ of $\worldsheet$,
and for each face $f \in T$ a local section
$\mathrm{or}_f  : U_f \to \worldsheetCoveringSpace$,
where $U_f$  is some open neighborhood of
$f$ in $\worldsheet$. By definition of the
integral of a density,
\begin{equation}
S_{\text{kin}}(\hat \stringmap) = \sum_{f
\in T} \int_{\mathrm{or}_f(f)}\mathcal{L}(\hat \stringmap)\label{wzwdenint}\text{.}
\end{equation}
One  immediately checks that this definition is independent of the choice
of the local sections: if one chooses for one face $f$ the other orientation, namely
$\worldsheetInvolution(\mathrm{or}_f)$, the
corresponding term in the sum (\ref{wzwdenint}), 
\begin{equation}
 \int_{\worldsheetInvolution(\mathrm{or}_f(f))}\mathcal{L}(\hat \stringmap) = - \int_{\mathrm{or}_f(f)}\worldsheetInvolution^{*}\mathcal{L}(\hat \stringmap)=  \int_{\mathrm{or}_f(f)}\mathcal{L}(\hat \stringmap)\text{,}
\end{equation}
gives the same contribution. It is also independent of
the choice of the triangulation. Furthermore,
if $\worldsheet$ is orientable, we can choose
a triangulation with a single face $f = \worldsheet$
and  get $S_{\text{kin}}(\hat \stringmap) = S_{\text{kin}}( \stringmap)$, which was
precisely our requirement on $S_{\text{kin}}(\hat \stringmap)$.

\vspace{3mm}  

We have already discussed in section \ref{ss_hol}
how to define  surface holonomies for an arbitrary
closed surface $\worldsheet$ with a map $\hat\stringmap \in
C^{\infty}(\worldsheetCoveringSpace, \targetspace)^{\orbifoldGroupElement,\worldsheetInvolution}$:
we have to choose a Jandl structure $\jandl$
on $\gerbe$. Then
$\mathrm{hol}_{\gerbe,\jandl}(\hat\stringmap,
\worldsheet)$ is defined in Definition \ref{hol_bg_def}
in such a way that if $\worldsheet$ is orientable,
it coincides by Theorem \ref{hol_bg_red}
with $\mathrm{hol}_\gerbe(\stringmap,\worldsheet)$. Remember
that a necessary condition on the existence
of a Jandl structure $\mathcal{J}=(\orbifoldGroupElement,-,-)$
was that the gerbes $\orbifoldGroupElement^{*}\gerbe$
and $\gerbe^{*}$ are stably isomorphic. We
already have encountered this condition for the orientable
WZW  model, so that it does not come as
an additional restriction. This leads us to the following

\begin{definition}
An unoriented WZW model consists of a compact connected Lie
group $\targetspace$, which is equipped with
an $\mathrm{Ad}$-invariant  metric $g$ on its Lie-algebra and a bi-invariant
gerbe $\gerbe$ with Jandl structure $\jandl$,
whose action of $\Z_2$ on $\targetspace$
is a parity transformation map $\orbifoldGroupElement
\in P(\targetspace,g,\gerbe)$.  To a closed
conformal surface $\worldsheet$ and a map   $\hat \stringmap \in
C^{\infty}(\worldsheetCoveringSpace, \targetspace)^{\orbifoldGroupElement,\worldsheetInvolution}$
the amplitude
\begin{equation}
A^\text{unor}_{g,\gerbe,\jandl}(\hat \stringmap, \worldsheet):=\exp \left( \im S_{\text{kin}}(\hat
\stringmap) \right) \cdot \mathrm{hol}_{\gerbe, \jandl}(\hat\stringmap,
 \worldsheet)\text{.}
\end{equation}
is assigned. 
\end{definition} 

According to the definition of both factors,
if $\worldsheet$ is orientable, we have
\begin{equation}
A^\text{unor}_{g,\gerbe,\jandl}(\hat \stringmap, \worldsheet) = A^\text{orble}_{g,\gerbe}(\hat \stringmap, \worldsheet)\text{.} 
\end{equation}
If $\worldsheet$ is even oriented, by equation
(\ref{wzwamplor}) we have
\begin{equation}
A^\text{unor}_{g,\gerbe,\jandl}(\hat \stringmap, \worldsheet)= A^\text{ortd}_{g,\gerbe}(\stringmap, \worldsheet) \text{.}
\end{equation}

\subsection{Crosscaps and the trivial
line bundle}
\label{ss_crosscaps}

In the following two sections we use the classification
of Jandl structures to classify unoriented
WZW models with a fixed gerbe $\gerbe$ and
a fixed parity transformation map $\orbifoldGroupElement\in
P(\targetspace,g,\gerbe)$. By Theorem \ref{bg_js_to}, the
set of equivalence classes of Jandl structures
of  $\gerbe$ with
the action of $\orbifoldGroup = \Z_2$
on $\targetspace$ defined by $\orbifoldGroupElement$ is a torsor over the flat
$\orbifoldGroup$-equivariant Picard group
$\mathrm{Pic}_0^\orbifoldGroup(\targetspace)$.
In this section we discuss a special element
of this group. 

On any manifold, there is the trivial line bundle $L_1 := \targetspace
\times \C$ with the trivial hermitian
metric and the trivial connection,
which is flat. It represents
the unit
element of the flat Picard group $\mathrm{Pic}_0(\targetspace)$.

Recall the following facts concerning
equivariant line bundles \cite{gomi1}.
There are two obstructions for
a given  line bundle 
 to admit equivariant structures:
the first depends on the bundle
and the group action,
namely that \begin{equation}\orbifoldGroupElement^{*}L \otimes L^{*} \cong L_1\label{equivobs}\text{,}\end{equation} which is still to be understood as an equation
of  hermitian
line bundles with connection. The
second obstruction is a class
in the group cohomology group $\mathrm{H}^2_{\text{Grp}}(\orbifoldGroup,U(1))$.
Now, if both obstructions are absent,
the possible equivariant structures
are parameterized by the group cohomology
group $\mathrm{H}^1_{\text{Grp}}(\orbifoldGroup,U(1))$
which is just the group of one-dimensional
characters of $\orbifoldGroup$.
In our case $\orbifoldGroup=\Z_2$
we have
\begin{eqnarray}
\mathrm{H}^1_{\text{Grp}}(\orbifoldGroup,U(1))
&=& \Z_2 \\
\mathrm{H}^2_{\text{Grp}}(\orbifoldGroup,U(1))
&=&0
\end{eqnarray}
so that the second obstruction
vanishes, and every  line
bundle $L$, which
satisfies the remaining  obstruction
(\ref{equivobs}) admits exactly
two $\orbifoldGroup$-equivariant structures. 

In particular $L_1$
itself satisfies (\ref{equivobs}).
We exhibit its two equivariant
structures explicitly. Remember from
section \ref{ss_js}, that we have
to choose an isomorphism
\begin{equation}
\varphi: \orbifoldGroupElement^{*}L_1
\to L_1
\end{equation} 
of  line bundles, such that $\varphi \circ \orbifoldGroupElement^{*}\varphi
= \Id_{L_1}$. So the both choices
are either $\varphi_1 = \Id_{\targetspace
\times \C}$ or $\varphi_{-1}: (x,z)
\mapsto (x,-z)$. We denote $L_1$
together with the equivariant structure
$\varphi_1$ by $L_1^{\orbifoldGroup}$. It represents
 the unit element of $\mathrm{Pic}_0^\orbifoldGroup(\targetspace)$.
We denote $L_1$ together with the
equivariant structure $\varphi_{-1}$
by $L_{-1}^{\orbifoldGroup}$. Note that $L_{-1}^{\orbifoldGroup}
\otimes L_{-1}^{\orbifoldGroup}
= L_{1}^{\orbifoldGroup}$ as equivariant
 line bundles.
Hence it
represents a non-trivial element
of order two in $\mathrm{Pic}_0^\orbifoldGroup(\targetspace)$.

The whole construction is completely
independent of $\targetspace$,
so $\mathrm{Pic}_0^\orbifoldGroup(\targetspace)$
always contains at least these
two elements. As a consequence, if
a gerbe $\gerbe$ admits a Jandl
structure $\jandl$, then $L_{-1}^{\orbifoldGroup}.\jandl$
is another, inequivalent Jandl
structure on $\gerbe$. We will now investigate
the difference between the corresponding unoriented WZW models.

We  work with local data, so
let $\lbrace \targetspaceCover_i \rbrace_{i \in I}$ be a good open
cover of $\targetspace$. Choose
all the sections that have been introduced
in section \ref{ss_ld}, and extract local
   data $(\jandllocij,
\jandlloci)$, $\jandlloccc$ of the
Jandl structure $\jandl$.  We also
explained how to extract a local
datum $\nu_i : \targetspaceCover_i
\to U(1)$ from an equivariant structure
on a  line bundle 
 over $\targetspace$.
The local datum of $L_{1}^{\orbifoldGroup}$
is the  constant global function $\nu_{1}=1$,
and the local datum of $L_{-1}^{\orbifoldGroup}$
is the  constant global function $\nu_{1}=-1$.

According to the definition of
the action of $\mathrm{Pic}_0^\orbifoldGroup(\targetspace)$
on $\mathrm{Jdl}(\gerbe,\orbifoldGroupElement)$,
the local data of $L_{-1}^{\orbifoldGroup}.\jandl$
are $(\jandllocij,\jandlloci)$
and $-\jandlloccc$.
Now observe the occurrences of the
local datum $\jandlloccc$ in  the
local holonomy formula (\ref{hol_ld}):
it appears for each orientation-reversing
vertex $v \in \bar V$. Following
our example in section \ref{ss_pp}, this happens
in the presence of a crosscap. We conclude
that the
amplitudes of both unoriented WZW
models with Jandl structures $\jandl$
and $L_{-1}^{\orbifoldGroup}.\jandl$
differ by a sign for each crosscap
in $\worldsheet$.

\subsection{Examples of target spaces}

\label{ss_examples}

We would like to discuss three
examples of target spaces, namely
the Lie groups $SU(2)$,
$SO(3)$, where the $\mathrm{Ad}$-invariant
metric
on their Lie algebras is given by their Killing forms, and the two-dimensional
torus $T^2=S^1\times S^1$ with the euclidean
scalar product.
The gerbes are  supposed to be  bi-invariant. 

\subsubsection{The Lie group ${SU}(2)$}

Following our general discussion, the actions
of $\Z_2$ on $SU(2)$ we have to
consider  are given by $\orbifoldGroupElement :g\mapsto g^{-1}$ and $\orbifoldGroupElement:g \mapsto -
g^{-1}$, where $-1\in Z(SU(2)) $ is the non-trivial element
in the center. The same maps were considered
in \cite{huiszoon1,brunner1,bachas1}.  

Fix a bi-invariant gerbe $\gerbe$ over $SU(2)$.
Up to stable isomorphism, this is $\gerbe
= \gerbe_0^{\otimes n}$, where $\gerbe_0$
is the basic gerbe over $SU(2)$ \cite{meinrenken1}.
By Lemma \ref{paritysu2}, both $\orbifoldGroupElement$'s
are parity transformation maps.  

The set $\mathrm{Jdl}(\gerbe,
\orbifoldGroupElement)$ is a torsor over $\mathrm{Pic}_0^{K}(SU(2))$
by Theorem \ref{bg_js_to}. In order
to compute  the group of equivariant
flat  line bundles, we first
observe
\begin{equation}
\mathrm{Pic}_0(\targetspace)=\mathrm{Hom}(\pi_1(\targetspace),
U(1))=0\text{,}
\end{equation}
since $SU(2)$ is simply connected.
So up to isomorphism there is only
one flat   line bundle,  the trivial one.  Hence there are exactly
two inequivalent Jandl structures
for each map $\orbifoldGroupElement$
and each bi-invariant gerbe $\gerbe$; this is
in agreement with the results of \cite{pradisi3,pradisi2}

\subsubsection{The Lie group $SO(3)$}

The center of $SO(3)$ is trivial,
so that we have only one action to consider,
namely by $\orbifoldGroupElement:
g \mapsto g^{-1}$.  Let $\mathcal{G}$
be a bi-invariant gerbe over $SO(3)$, such
that $\orbifoldGroupElement^{*}\gerbe$ and
$\gerbe^{*}$ are stably isomorphic. Such
gerbes for example are constructed up to stable isomorphism in
\cite{gawedzki2}. We have
 to investigate the group $\mathrm{Pic}_0^{K}(SO(3))$ of
flat equivariant  line
bundles. Again we first consider
the group $\mathrm{Pic}_0(SO(3))$
of flat  line bundles
and classify equivariant structures
on them.  

By $\pi _{1}(SO(3)) = \Z_2$ we
have
\begin{equation}
\mathrm{Hom}(\pi_{1}(SO(3)) ,U(1)) =\mathrm{Hom}(\Z_{2},U(1)) = \Z_{2}\text{,}
\end{equation}
so there are - up to isomorphism
- two flat  line bundles.
We will give them explicitly: As
$SO(3)$ is the quotient of $SU(2)$
by $q:g \mapsto -g$, the two flat
line bundles over $SO(3)$ correspond
to the two equivariant  flat line bundles   over $SU(2)$, namely $L_{1}
  ^{\orbifoldGroup}$ and $L_{-1}^{\orbifoldGroup}$
 . 

Clearly, $L_{1}^{\orbifoldGroup}$
descends to the trivial flat 
line bundle $\tilde L_1\to
SO(3)$, which admits equivariant
structures,  more precisely, according to
the discussion in section \ref{ss_crosscaps}, there are two of them. $L_{-1}^{\orbifoldGroup}$ descends
to a non-trivial flat 
line bundle $\tilde L_{-1}\to  SO(3)$, and we have to ask whether
it admits equivariant structures,
which is equivalent to the condition,
that
\begin{equation}
\exteriorDifferential \tilde L_{-1}:=\orbifoldGroupElement^{*}\tilde
L_{-1} \otimes \tilde L_{-1}^{*}
\cong \tilde L_{1}\text{.}
\label{so3equiv}\end{equation}
Now $\exteriorDifferential \tilde L_{-1}$ is a flat  line bundle, and hence either isomorphic
to $\tilde L_{-1}$ or to $\tilde L_{1}$. Because $\mathrm{Pic}_0(SO(3))$
 is a group of order two, we have
 $\tilde L_{-1} \otimes \tilde L_{-1} = \tilde L_{1}$. The assumption
$\exteriorDifferential \tilde L_{-1}
\cong \tilde L_{-1}$ would therefore mean $\orbifoldGroupElement^{*}\tilde
L_{-1} \cong  \tilde L_{1}$ which
is a contradiction since $\tilde L_{1}$ is the trivial bundle and
$\orbifoldGroupElement^{*}\tilde
L_{-1}$ is not. Hence (\ref{so3equiv})
is true, and $\tilde L_{-1}^{*}$
admits two equivariant structures.

All together, there are four equivariant
flat  line bundles over
$SO(3)$ and hence four Jandl structures
on $\gerbe$; again, this is in agreement
with \cite{pradisi3,pradisi2}. 
 
\subsubsection{The two-dimensional Torus
$T^2$}

For dimensional reasons, all gerbes
over $T^2$ are trivial and have
 curvature $H=0$. This allows us to discuss
 an example with a parity transformation
 map $\orbifoldGroupElement$, which is not
 of the form $\orbifoldGroupElement = l_z
 \circ \mathrm{Inv}$ but simply the identity
 map $\orbifoldGroupElement = \id$. This
 allows us to make contact with \cite{bianchi1}.  

Now let $\gerbe$ be a bi-invariant gerbe over $T^2$.   The set $\mathrm{Jdl}(\gerbe,
\Id)$ is a torsor over $\mathrm{Pic}_0^{K}(T^2)$
by Theorem \ref{bg_js_to}, which
is isomorphic to $\mathrm{H}^1_K(T^2,U(1))$ by equation
(\ref{bg_js_pic}). The Borel space associated to the trivial $\orbifoldGroup$-action
is $T^2_K = E\Z_2 \times T^2$.
With $E\Z_2=\R P^\infty$
we have
\begin{eqnarray*}
\mathrm{H}^1_K(T^2,U(1)) &=& H^1(T_\orbifoldGroup^2,U(1))
\\ &=&\mathrm{H}^1(\R
P^\infty,U(1)) \oplus \mathrm{H}^1(T^2,U(1))
\\ &=&\Z_2 \oplus U(1) \oplus U(1)
\\ &=&\Z_2 \oplus T^2
\text{.}
\end{eqnarray*}

We now assume that the gerbe $\gerbe$
admits a Jandl structure $\mathcal{J}=(\Id,\mathcal{A},\varphi)$.
In particular, $\mathcal{A}=(A,\alpha)$
is a stable isomorphism from $\gerbe$
to $\gerbe^*$.
Recall that a gerbe $\gerbe$ consist of the following data:
a surjective submersion $\pi:Y \to \targetspace$,
a  line bundle $L \to Y^{[2]}$,  an isomorphism $\mu$, and a
2-form $C \in \Omega^2(Y)$. Recall further
that here  $A$ is a  line bundle
  over $Z=Y^{[2]}$, and both projections $p$ and $p'$ from $Z$ to $Y$ coincide with
$\pi_2, \pi_1:Y^{[2]} \to Y$.

 The condition on the curvature
of $A$ in Definition \ref{bg_si_df}
now reads
\begin{equation}
\mathrm{curv}(A) = \pi_1^*C + \pi_2^*C\text{.} \end{equation} 
Furthermore, since for all gerbes
the curving $C$ satisfies
$-\pi_{2}^*C+\pi_{1}^*C=\mathrm{curv}(L)$,
we have
\begin{equation}
2\pi_2^{*}C =  \mathrm{curv}(A)-\mathrm{curv}(L)
\text{,}
\end{equation} 
which is an equation of 2-forms on $Y^{[2]}$. On the right hand side we have a closed 2-form which
defines an integral class in cohomology.
Since $\pi_2$ is a surjective submersion,
also $2C$ defines a  class in $\mathrm{H}^{2}(Y,\Z)$.

Because the gerbe $\mathcal{G}$ is trivial,
we can choose a trivialization $\mathcal{T}$ and obtain
the 2-form $B \in \Omega^2(\targetspace)$
as in Definition \ref{hol_bg_or}, which satisfies $\pi^{*}B=C+\mathrm{curv}(T)$
and $\exteriorDifferential B = H =
0$.  Usually one chooses $\mathcal{T}$ such that
$B$ is constant, then it is nothing but the Kalb-Ramond \textquotedblleft B-Field\textquotedblright. Because $\pi$ is  also
a surjective submersion it follows that  $2B$ defines a class
in $\mathrm{H}^2(\targetspace,\Z)$.
Thus we have derived the quantization condition
that the $B$-Field has half
integer valued periods. This condition was originally
found
in \cite{bianchi1}
by an analysis of the bulk spectrum of right and left movers. 

\newcommand{\etalchar}[1]{$^{#1}$}

\end{document}